\documentclass[
preprintnumbers,
nofootinbib,
 amsmath,amssymb,
 prd,aps,superscriptaddress,twocolumn,
floatfix,
]{revtex4-2}

\usepackage{aas_macros}
\usepackage{graphicx}
\usepackage{amsmath}
\usepackage[caption=false]{subfig}
\usepackage{siunitx}
\usepackage{placeins}
\usepackage{color}
\usepackage{standalone}
\usepackage{dcolumn}
\usepackage{tensor}
\usepackage{bm}
\usepackage{microtype}
\usepackage{etoolbox}
\usepackage{amssymb}
\usepackage{mathrsfs}
\usepackage{accents}
\usepackage[normalem]{ulem}
\usepackage{soul} 
\usepackage[dvipsnames]{xcolor}
\usepackage[colorlinks,urlcolor=NavyBlue,citecolor=NavyBlue,linkcolor=NavyBlue,pdfusetitle]{hyperref}
\usepackage[all]{hypcap}
\usepackage[inline]{enumitem}
\usepackage[utf8]{inputenc}
\usepackage{xspace}
\usepackage[printonlyused, nolist]{acronym}
\usepackage{lineno}

\newcommand{\Mtotal}{\ensuremath{M_{\rm tot}}\xspace}
\newcommand{\Mtotaltrue}{\ensuremath{\hat{M}_{\rm tot}}\xspace}
\newcommand{\msun}{\ensuremath{M_{\odot}}\xspace}

\newcommand{\zcritvalue}{\ensuremath{30}\xspace}
\newcommand{\zcrit}{\ensuremath{z_\mathrm{crit}}\xspace}
\newcommand{\fcrit}{\ensuremath{f^{\mathrm{PBH}}_{\mathrm{III}}}\xspace}

\newcommand{\aIII}{\ensuremath{a_{\rm III}}\xspace}
\newcommand{\bIII}{\ensuremath{b_{\rm III}}\xspace}
\newcommand{\zIII}{\ensuremath{z_{\rm III}}\xspace}
\newcommand{\npbh}{\ensuremath{\dot{n}_{\rm PBH}}\xspace}
\newcommand{\nIII}{\ensuremath{\dot{n}_{\rm III}}\xspace}
\newcommand{\ptot}{\ensuremath{p_{\rm tot}}\xspace}
\newcommand{\rhoPHM}{\ensuremath{\rho_{\rm PHM}}\xspace}
\newcommand{\rhoP}{\ensuremath{\rho_{\rm P}}\xspace}
\newcommand{\chieff}{\ensuremath{\chi_{\rm eff}}}
\newcommand{\Pp}{\ensuremath{P_\text{p}}\xspace}
\newcommand{\IMRXPHM}{\texttt{IMRPhenomXPHM}\xspace}
\newcommand{\IMRXP}{\texttt{IMRPhenomXP}\xspace}
\newcommand{\IMRP}{\texttt{IMRPhenomPv2}\xspace}

\newcommand{\penncosmos}{\affiliation{Institute for Gravitation and the Cosmos, Department of Physics, Pennsylvania State University, University Park, PA, 16802, USA}}

\newcommand{\pennastro}{\affiliation{Department of Astronomy \& Astrophysics, Pennsylvania State University, University Park, PA, 16802, USA}}
\newcommand{\cardiff}{\affiliation{School of Physics and Astronomy, Cardiff University, Cardiff, UK, CF24 3AA}
}
\newcommand{\sapienza}{Dipartimento di Fisica, Sapienza Università 
	di Roma, Piazzale Aldo Moro 5, 00185, Roma, Italy}
\newcommand{\infn}{INFN, Sezione di Roma, Piazzale Aldo Moro 2, 00185, Roma, Italy}

\graphicspath{{figures/}}

\begin{document}
\title{Measuring properties of primordial black hole mergers at cosmological distances: effect of higher order modes in gravitational waves}

\author{Ken K.~Y.~Ng}
\email{kng15@jhu.edu}
\affiliation{William H. Miller III Department of Physics and Astronomy, Johns Hopkins University, Baltimore, Maryland 21218, USA}
\affiliation{LIGO, Massachusetts Institute of Technology, Cambridge, Massachusetts 02139, USA}
\affiliation{Kavli Institute for Astrophysics and Space Research, Massachusetts Institute of Technology, Cambridge, Massachusetts 02139, USA}
\author{Boris Goncharov}
\affiliation{Gran Sasso Science Institute (GSSI), I-67100 L'Aquila, Italy}
\affiliation{INFN, Laboratori Nazionali del Gran Sasso, I-67100 Assergi, Italy}
\author{Shiqi Chen}
\affiliation{LIGO, Massachusetts Institute of Technology, Cambridge, Massachusetts 02139, USA}
\affiliation{Kavli Institute for Astrophysics and Space Research, Massachusetts Institute of Technology, Cambridge, Massachusetts 02139, USA}
\author{Ssohrab Borhanian}
\penncosmos
\affiliation{Theoretisch-Physikalisches Institut, Friedrich-Schiller-Universit\"at Jena, 07743, Jena, Germany}
\author{Ulyana Dupletsa}
\affiliation{Gran Sasso Science Institute (GSSI), I-67100 L'Aquila, Italy}
\affiliation{INFN, Laboratori Nazionali del Gran Sasso, I-67100 Assergi, Italy}
\author{Gabriele~Franciolini}
\affiliation{\sapienza}
\affiliation{\infn}
\author{Marica Branchesi}
\affiliation{Gran Sasso Science Institute (GSSI), I-67100 L'Aquila, Italy}
\affiliation{INFN, Laboratori Nazionali del Gran Sasso, I-67100 Assergi, Italy}
\author{Jan Harms}
\affiliation{Gran Sasso Science Institute (GSSI), I-67100 L'Aquila, Italy}
\affiliation{INFN, Laboratori Nazionali del Gran Sasso, I-67100 Assergi, Italy}
\author{Michele Maggiore}
\affiliation{D\'epartement de Physique Th\'eorique, Universit\'e de Gen\`eve, 24 quai E. Ansermet, CH-1211 Geneva, Switzerland}
\affiliation{Gravitational Wave Science Center (GWSC), Universit\'e de Gen\`eve, CH-1211 Geneva, Switzerland}
\author{Antonio~Riotto}
\affiliation{D\'epartement de Physique Th\'eorique, Universit\'e de Gen\`eve, 24 quai E. Ansermet, CH-1211 Geneva, Switzerland}
\affiliation{Gravitational Wave Science Center (GWSC), Universit\'e de Gen\`eve, CH-1211 Geneva, Switzerland}

\author{B. S. Sathyaprakash}
\penncosmos \pennastro \cardiff
\author{Salvatore Vitale}
\affiliation{LIGO, Massachusetts Institute of Technology, Cambridge, Massachusetts 02139, USA}
\affiliation{Kavli Institute for Astrophysics and Space Research, Massachusetts Institute of Technology, Cambridge, Massachusetts 02139, USA}

\date{\today}

\begin{abstract}
Primordial black holes (PBHs) may form from the collapse of matter overdensities shortly after the Big Bang.
One may identify their existence by observing gravitational wave (GW) emissions from merging PBH binaries at high redshifts $z\gtrsim 30$, where astrophysical binary black holes (BBHs) are unlikely to merge.
The next-generation ground-based GW detectors, Cosmic Explorer and Einstein Telescope, will be able to observe BBHs with total masses of $\mathcal{O}(10-100)~\msun$ at such redshifts.
This paper serves as a companion paper of Ref.~\citep{Ng:2021sqn}, focusing on the effect of higher-order modes (HoMs) in the waveform modeling, which may be detectable for these high redshift BBHs, on the estimation of source parameters.
We perform Bayesian parameter estimation to obtain the measurement uncertainties with and without HoM modeling in the waveform for sources with different total masses, mass ratios, orbital inclinations and redshifts observed by a network of next-generation GW detectors.
We show that including HoMs in the waveform model reduces the uncertainties of redshifts and masses by up to a factor of two, depending on the exact source parameters.
We then discuss the implications for identifying PBHs with the improved single-event measurements, and expand the investigation of the model dependence of the relative abundance between the BBH mergers originating from the first stars and the primordial BBH mergers as shown in Ref.~\citep{Ng:2021sqn}.
\end{abstract}

\maketitle

\preprint{ET-0231A-22, CE-P2200006}

\section{Introduction}

An interesting possibility is that a fraction of the merger events detected by the LIGO-Virgo-KAGRA (LVK) Collaboration may be due to primordial BHs~(PBHs)~\cite{Zeldovich:1967lct,Hawking:1974rv,Chapline:1975ojl,Carr:1975qj} formed from the collapse of sizable overdensities in the radiation-dominated early universe~\cite{Ivanov:1994pa,GarciaBellido:1996qt,Ivanov:1997ia,Blinnikov:2016bxu}. In this scenario, PBHs are not clustered at formation \cite{Ali-Haimoud:2018dau,Desjacques:2018wuu,Ballesteros:2018swv,MoradinezhadDizgah:2019wjf,Inman:2019wvr,DeLuca:2020jug}, they are born spinless~\cite{DeLuca:2019buf, Mirbabayi:2019uph} and may assemble in binaries via gravitational decoupling from the Hubble flow before the matter-radiation equality \cite{Nakamura:1997sm,Ioka:1998nz} (see~\cite{Polnarev:1985btg,Khlopov:2008qy,Sasaki:2018dmp,Green:2020jor,Franciolini:2021nvv} for reviews).
After their formation, PBH binaries may be affected by a phase of baryonic mass accretion at redshifts smaller than $z \sim 30$, which would modify the PBH masses, spins and merger rate~\cite{DeLuca:2020bjf, DeLuca:2020qqa}.

Analysing the population properties of masses, spins, and redshifts of binary black holes (BBHs) in the LVK's second catalog~\citep{GWTC2}, several studies constrained the potential contribution from PBH binaries to current data~\citep{DeLuca:2020sae,Wong:2020yig,DeLuca:2020qqa,Hutsi:2020sol,Hall:2020daa,DeLuca:2021wjr,Franciolini:2021tla,Mukherjee:2021ags,2022arXiv220905959F}.
However, these analyses require precise knowledge of the astrophysical BBH ``foreground'' in order to verify if there is a PBH subpopulation within the BBHs observed at low redshifts~\citep{Franciolini:2021tla,Franciolini:2021xbq}.
Such analyses are limited by the horizon of current GW detectors, $z\lesssim 3$ at their design sensitivity~\citep{Hall:2019xmm}, and are subject to significant uncertainties on the mechanisms of BBH formation in different astrophysical environments, such as galactic fields~\citep{OShaughnessy:2016nny,Dominik:2012kk,Dominik:2013tma,Dominik:2014yma,deMink:2015yea,Belczynski:2016obo,Stevenson:2017tfq,Mapelli:2019bnp,Breivik:2019lmt,Bavera:2019fkg,Broekgaarden:2019qnw}, dense star clusters~\citep{2000ApJ...528L..17P,Antonini:2020xnd,Santoliquido:2020axb,Rodriguez:2015oxa,Rodriguez:2016kxx,Rodriguez:2018rmd,DiCarlo:2019pmf,Kremer:2020wtp,Rodriguez:2015oxa,Rodriguez:2016kxx,Rodriguez:2018rmd,Antonini:2020xnd}, active galactic nuclei~\citep{Bartos:2016dgn,Yi:2019rwo,Yang:2019cbr,Yang:2020lhq,Grobner:2020drr,Tagawa:2019osr,Tagawa:2020qll,Tagawa:2020dxe,Samsing:2020tda}, or from the collapse of Population III (Pop III) stars~\citep{Kinugawa:2014zha,Kinugawa:2015nla,Hartwig:2016nde,Belczynski:2016ieo}.

Instead, searching for PBHs at high redshifts where astrophysical BHs have not merged yet may mitigate most of the issues caused by the astrophysical foreground.
The PBH merger rate increases with redshift~\citep{Raidal:2018bbj}, while the astrophysical contribution decreases at $z \lesssim 30$~\citep{Belczynski:2016ieo,Tanikawa:2021qqi,Hijikawa:2021hrf}.
The proposed next-generation detectors, such as the Cosmic Explorer (CE)~\citep{Evans:2016mbw,Reitze:2019iox,CEHS} and the Einstein Telescope (ET)~\citep{Punturo:2010zz,Maggiore:2019uih}, whose horizons are up to $z\sim 100$ for stellar-mass BBHs~\cite{Hall:2019xmm,Iacovelli:2022bbs}, may provide a unique opportunity to test and shed light on the primordial origin of BH mergers at high redshifts.
A key question is therefore to understand the uncertainties related to the measurements of the source parameters, such as the redshift, masses and spins.

In Ref.~\citep{Ng:2021sqn}, we established the possibility of identifying the PBH mergers with masses of 20 and 40~\msun at $z\geq 40$ using single-event redshift measurements.
We also discussed how the prior knowledge of relative abundance between Pop~III and PBH mergers affects the statistical significance, assuming that there is a critical redshift, $\zcrit=\zcritvalue$, above which no astrophysical BBHs are expected to merge.
The results were based on full Bayesian parameter estimation with a waveform model, \IMRXPHM, which includes the effects of spin precession and higher-order modes (HoMs)~\citep{Pratten:2020ceb,Pratten:2020fqn,Garcia-Quiros:2020qpx}.
In this paper, we show the importance of HoMs to the parameter estimation of the high redshift BBHs at $z\geq10$ in the context of PBH detections.
We compare the Bayesian posteriors of the relevant parameters obtained by \IMRXPHM and the similar waveform family without HoMs, \IMRP~\cite{Hannam:2013oca,Husa:2015iqa,Khan:2015jqa} to systematically study the improvement on measurements due to the HoM modeling in the waveform.

We first recap the details of our simulations and the settings of the parameter estimation in Sec.~\ref{sec:simulations}.
Then, we show whether and how \IMRXPHM performs better when measuring redshift (Sec.~\ref{sec:redshift}), as well as masses and spins (Sec.~\ref{sec:intrinsic}), for BBHs with different sets of the source-frame total mass, mass ratio, orbital inclination, and redshift.
Finally, in Sec.~\ref{sec:implications}, we re-examine the estimation of the probability that a single source originated from PBHs using redshift measurements under different choices of $\zcrit$, and discuss the possible implications of the mass and spin measurements for PBH detections.

\section{Simulation details}\label{sec:simulations}
As in Ref.~\citep{Ng:2021sqn}, we simulate BBHs at five different redshifts, $\hat{z}=10,$ 20, 30, 40 and 50.
The hat symbol denotes the true value of a parameter here and throughout the paper.
To encompass the detectable mass range, we choose the total masses in the source frame to be $\Mtotaltrue=5$, 10, 20, 40, and 80 $\msun$, with mass ratios $\hat{q}=1$, 2, 3, 4 and 5.
Here, we define $q \equiv m_1/m_2$ for $m_1>m_2$, where $m_1$ and $m_2$ are the primary and secondary mass, respectively.
For each mass pair, we further choose four orbital inclination angles, $\hat{\iota}=0$ (face-on), $\pi/6$, $\pi/3$, and $\pi/2$ (edge-on).
All simulated BBHs are non-spinning, as we expect that PBHs are born with negligible spins~\citep{Bianchi:2018ula, Mirbabayi:2019uph,DeLuca:2019buf} and may be spun-up by accreting materials at later times~\citep{Bianchi:2018ula, DeLuca:2020bjf, DeLuca:2020fpg, DeLuca:2020qqa}.
However, we do \emph{not} assume zero spins when performing parameter estimation of the source parameters and instead allow for generic spin-precession.
For each of these 500 sources, the sky location and polarization angle are chosen to maximize the signal-to-noise ratio (SNR) for each source.
The reference orbital phase and GPS time are fixed at $0$ and $1577491218$, respectively.
The baseline detector network is a 40-km CE in the United States, 20-km CE in Australia, and ET in Europe.
We only analyze simulated sources whose network SNRs are larger than 12.
We use Planck 2018 Cosmology when calculating the luminosity distance $d_L$ at a given redshift~\cite{Planck:2018vyg}.

We employ a nested sampling algorithm~\cite{Skilling:2006gxv,2020MNRAS.493.3132S} packaged in \textsc{Bilby}~\citep{Ashton:2018jfp} to obtain posterior probability densities.
As we are only interested in the uncertainty caused by the loudness of the signal, we use a zero-noise realization~\citep{Vallisneri:2007ev} for the Bayesian inference and mitigate the offsets potentially caused by Gaussian fluctuations~\citep{Rodriguez:2013oaa}.
To ensure our results are free from the systematics due to the difference in the two waveform families, we use the same waveform family for both simulating the waveforms and calculating the likelihood.
That is, we use the \IMRXPHM (\IMRP) waveform \textit{template} to analyze the \IMRXPHM (\IMRP) simulated waveforms~\footnote{See Ref.~\citep{Purrer:2019jcp} for the analysis of waveform systematics for the next-generation GW detectors.}.
The low-frequency cut off in the likelihood calculations is 5~Hz for all sources.

As in Ref.~\citep{Ng:2021sqn}, we first sample the parameter space with uniform priors on the detector-frame total mass, $\Mtotal^{D} = \Mtotal(1+z)$, between $[0.5,1.5]\Mtotaltrue^{D}$, and $q$ between $[1,10]$.
The prior on redshift is uniform in the comoving rate density, $\propto \frac{dV_c}{dz}\frac{1}{1+z}$, between $[z(\hat{d}_L/10), z(5\hat{d}_L)]$.
In Sec.~\ref{sec:implications}, we will revisit the physically motivated prior on redshift.
We use uniform priors for other parameters: the sky position, the polarization angle, the orbital inclination, the spin orientations, the spin magnitudes, the arrival time and the phase of the signal at the time of arrival.

Then, we reweigh the posteriors into uniform prior on the source-frame primary mass, $m_1$, and the inverse mass ratio $1/q$ (which is between $[0.1,1]$).
Strictly speaking, the marginalized one-dimensional priors on $m_1$ and $1/q$ are not exactly uniform after the reweighing because the boundary of the square domain of $(\Mtotal^{D}, q)$ transforms into a different shape according to the Jacobian.
For example, the marginalized prior on the redshift and that on the inverse mass ratio have additional factors of $1/(1+z)$ and $q/(q+1)$, respectively, upon the coordinate transformation.
However, we find that such boundary effect has negligible effect on the posteriors.
As we will discuss below, the degeneracy among different parameters and the scaling in $p_0(z)$ is more significant.

\section{Redshift measurement in presence of higher-order modes}\label{sec:redshift}
\begin{figure}[!htb]
    \centering
    \subfloat{\includegraphics[width=0.925\columnwidth]{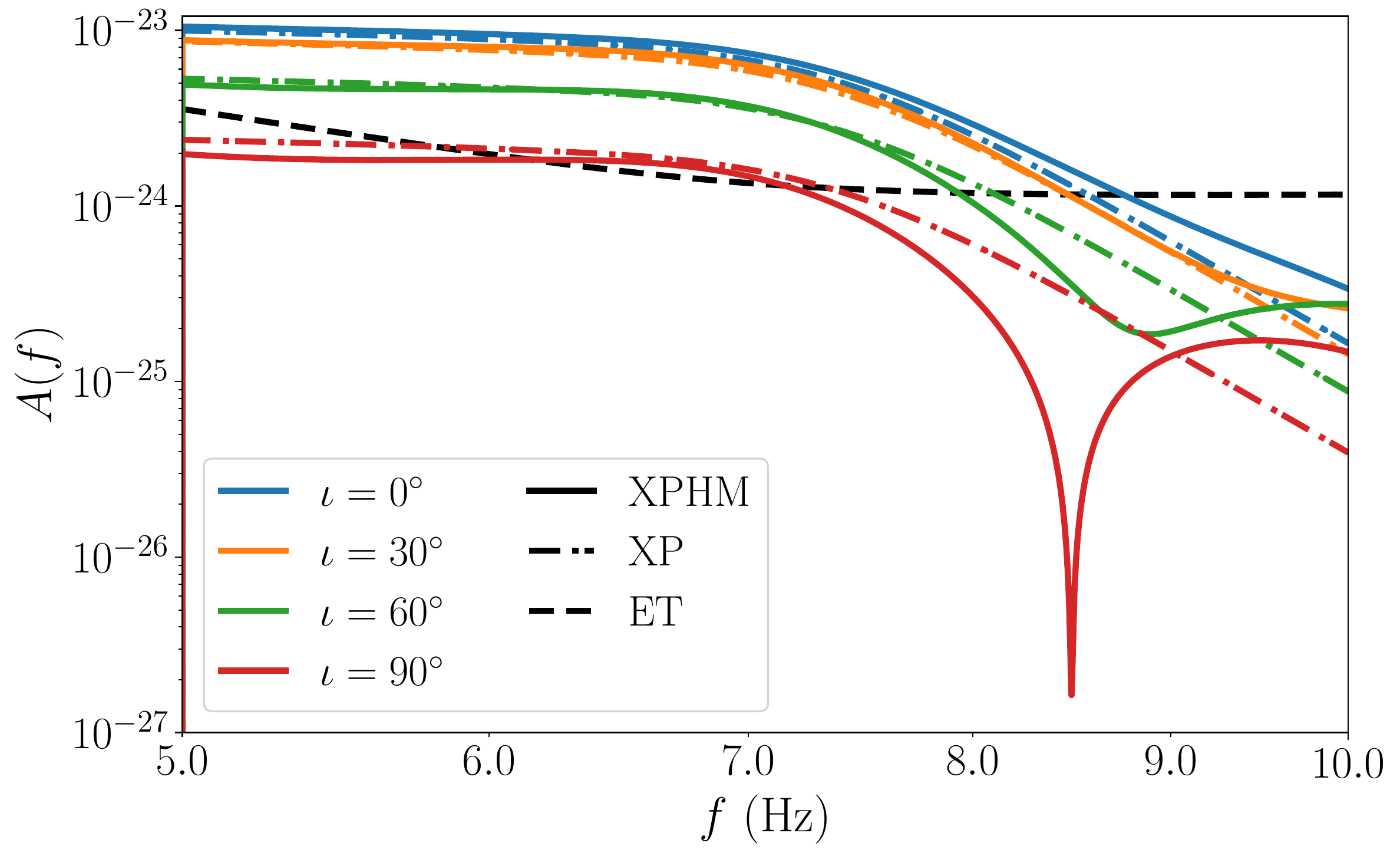}}
    \quad
    \subfloat{\hspace*{0.25cm}\includegraphics[width=0.9\columnwidth]{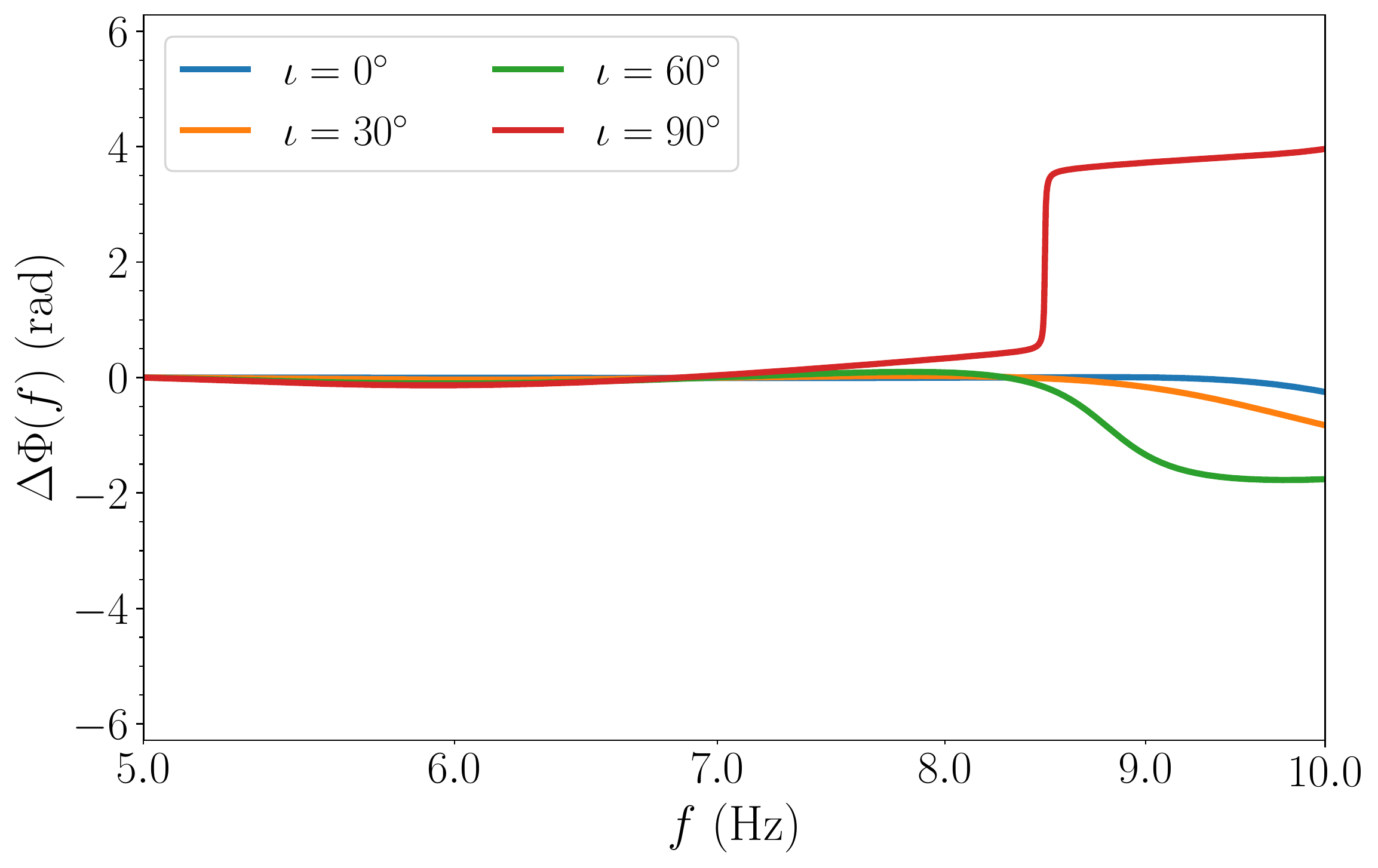}}
    \caption{Comparison between waveforms without HoMs (\IMRXP) and with HoMs (\IMRXP) for BBHs with $(\Mtotaltrue, \hat{z}, \hat{q})=(80~\msun, 30, 1)$ and $\hat{\iota}=0^{\circ}, 30^{\circ}, 60^{\circ}$ and $90^{\circ}$ (blue, orange, green and red, respectively).
    \textit{Top panel}: Strain amplitudes for \IMRXPHM (solid lines) and \IMRXP (dotted lines) projected on ET's detector frame.
    \textit{Bottom panel}: Phase difference between \IMRXPHM and \IMRXP at each frequency.
    In all systems, the right ascension angle, declination angle and polarization angle are $110^{\circ}$, $45^{\circ}$ and $93^{\circ}$, respectively.}
    \label{fig.wfs}
\end{figure}

Since each HoM has a different angular emission spectrum, including HoMs in the waveform model breaks the distance-inclination degeneracy characteristic of the dominant $(2, 2)$ harmonic mode~\cite{Usman:2018imj,Chen:2018omi}.
The interference of additional HoMs can result in amplitude modulation, similar to what can be induced by spin precession~\cite{Apostolatos:1994mx,Garcia-Quiros:2020qpx,Pratten:2020fqn}.
For example, in the top panel of Fig.~\ref{fig.wfs} we show the Fourier amplitude of a BBH with $(\Mtotaltrue, \hat{z}, \hat{q})=(80~\msun, 30, 1)$ and $\hat{\iota}=0^{\circ}, 30^{\circ}, 60^{\circ}$ and $90^{\circ}$ (blue, orange, green and red, respectively).
To reduce the systematics between waveform families due to differences in precessing frame mapping, we compare \IMRXPHM (solid lines) and \IMRXP~\cite{Pratten:2020ceb} (dotted lines) instead.
The amplitude modulation of the waveforms with HoMs -- which is stronger for inclination angles close to $90^\circ$ -- is apparent and helps improving the estimation of the distance and the inclination.
By contrast, for the waveforms without HoMs the main effect of increasing the inclination angle is to reduce the Fourier amplitude, which qualitatively shows why the two parameters are partially degenerate when only the (2,2) mode is used\footnote{The degeneracy is worst at small inclination angles, see Sec.~\ref{sec:iota}.}.
The other contribution is the phase modulation in the later part of the waveform due to HoMs.
To visualize this effect, we show the phase difference between \IMRXP and \IMRXPHM, $\Delta\Phi(f)$, in the bottom panel of Fig.~\ref{fig.wfs}.
Whereas the (2,2) mode of the inspiral defines the waveform up to $\approx 8$~Hz, after which the ringdown takes over, HoMs of the inspiral extend to higher frequencies.
The interference of the HoMs and the (2,2) ringdown piles up a significant phase modulation, and hence improves the measurement of inclination.

Moreover, the parameters, $q$, $\iota$ and $\Mtotal$, determine the amplitude of each mode.
The uncertainties of distance (and hence redshift) are thus sensitive to the values of $(q, \iota, \Mtotal)$ with other parameters fixed.
In this section, we will quantify the variation of the redshift uncertainty due to each intrinsic parameter one at a time.
We will also show which region of redshift gain the most from the presence of HoMs in the waveform model.
In the following figures, blue (red) violins represent the posteriors obtained by \IMRXPHM (\IMRP).

\subsection{Orbital inclination}\label{sec:iota}
We first discuss the role of orbital inclination in the redshift measurements using HoM waveforms.
Waveform models which only contain (2, 2) mode suffer from the distance-inclination degeneracy of the mode, especially for nearly face-on $(\iota \simeq 0)$ systems whose amplitude scales as $\sim (1-\iota^2/2)/d_L$.
On the other hand, each HoM corresponds to spherical harmonics with a different angular response as a function of $\iota$.
If the waveform models are sensitive to HoMs, measuring the relative amplitudes of the HoMs provides better constraints on the orbital inclination angle, and thus reduces the distance-inclination degeneracy.
\begin{figure}[!htb]
    \centering
    \includegraphics[width=0.9\columnwidth]{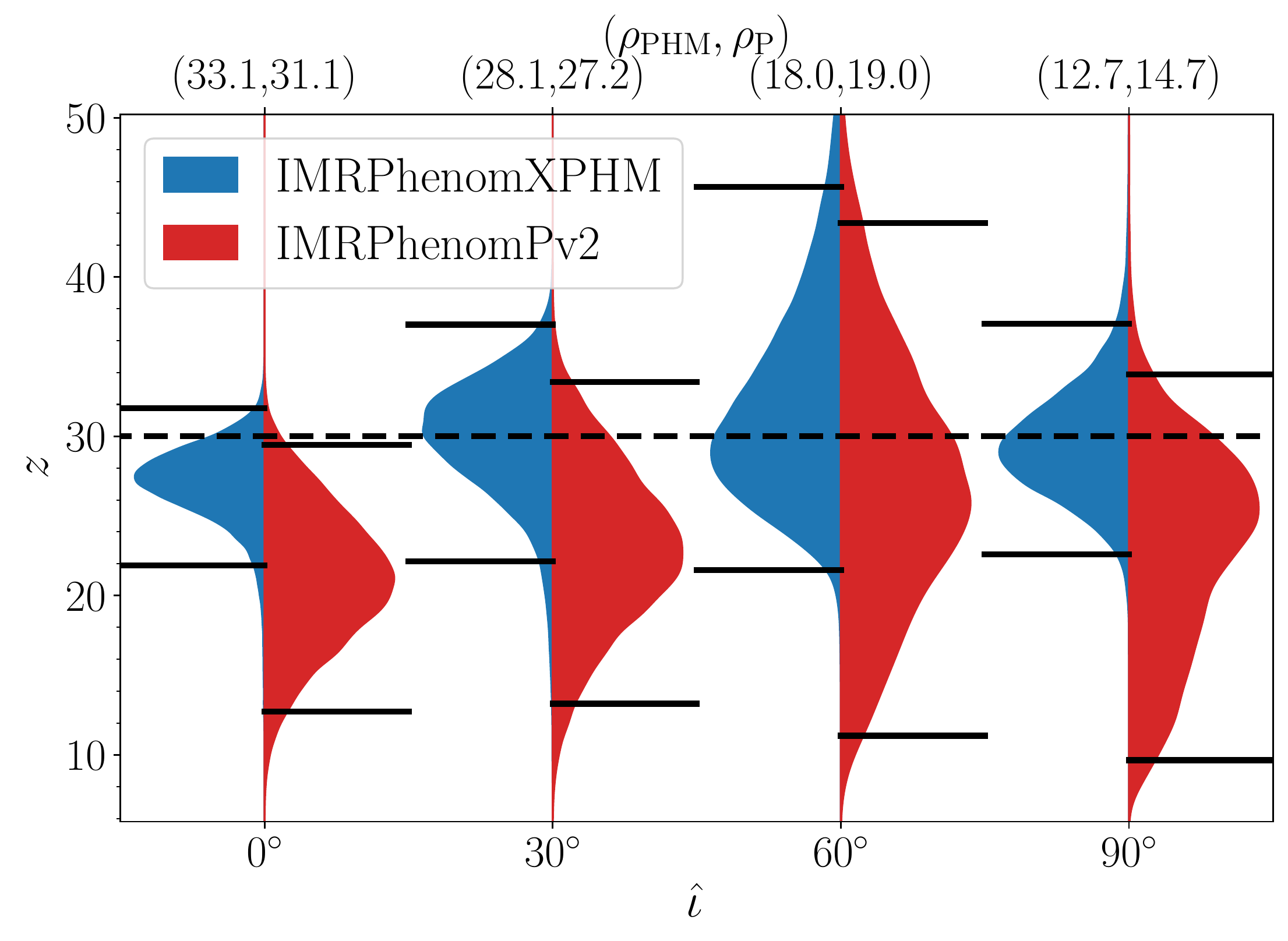}
    \caption{Posteriors of redshift for sources with $(\Mtotaltrue,\hat{z},\hat{q})=(40\msun,30,1)$ at $\hat{\iota}=0^{\circ}, 30^{\circ}, 60^{\circ}$ and $90^{\circ}$, obtained with HoM (blue, \IMRXPHM) and without HoM (red, \IMRP). The solid horizontal lines show the 95\% credible intervals, whereas the dashed lines mark $\hat{z}$. Top axis shows the optimal SNR of \IMRXPHM waveform, $\rhoPHM$, and that of \IMRP waveform, $\rhoP$. The detector network is CE-CES20-ET.
    We also note that the SNR does not necessarily increase with the addition of HoM because they may interfere destructively at large values of $\hat{\iota}$.}
    \label{fig:zposterior_iota}
\end{figure}

We now quantify the improvement on the redshift measurements due to the presence of HoMs with varying inclination angles.
In Fig.~\ref{fig:zposterior_iota}, we show the redshift posteriors obtained by the two waveform models for sources with $(\Mtotaltrue,\hat{z},\hat{q})=(40\msun,30,1)$ at $\hat{\iota}=0^{\circ}, 30^{\circ}, 60^{\circ}$ and $90^{\circ}$.
Indeed, the redshift uncertainties are generally smaller in the cases of \IMRXPHM than those of \IMRP.
The decrease in the uncertainties is about $\sim 30\%-50\%$.

Notably, the lower bound of redshift uncertainties increases from $z\sim 10$ in the cases of \IMRP to $z\sim 20$ in the cases of \IMRXPHM.
This improvement due to HoMs is particularly interesting for determining the astrophysical or primordial origin of BBHs.
If the redshift measurement of a system is precise enough to rule out the epoch of the astrophysical BBHs, one may even use a single measurement to identify the existence of primordial BBH, as discussed in Ref.~\cite{Ng:2021sqn}.
On the other hand, even if a single measurement is not conclusive enough, the improvement on redshift measurements reduces the required number of events to conduct a statistical analysis over a population of high-redshift BBHs~\citep{Ng:2020qpk,Ng:2022agi}.

As $\hat{\iota}$ increases, the SNR decreases with the amplitude of the dominant $(2,2)$ harmonic.
One might expect the redshift uncertainty would then increase with $\hat{\iota}$.
Indeed, as shown in Fig.~\ref{fig:zposterior_iota}, the uncertainty raises from $\hat{\iota}=0^{\circ}$ to $\hat{\iota}=60^{\circ}$ but shrinks when the system is edge on $(\hat{\iota}=90^{\circ})$.
For the edge-on system, the $\times$-polarization content is very sensitive to small changes of $\iota$, which makes it possible to obtain better estimates of $\iota$ and to break the distance-inclination degeneracy. The other feature worth-mentioning is the apparent bias in face-on systems.
As discussed in Ref.~\cite{Ng:2021sqn}, this is caused by the physical cut-off of the parameter space: an overestimation of the redshift cannot be compensated by the increase of $\cos \iota$ beyond 1.

We also note that the HoM-improvements and features of the redshift uncertainties discussed above are similar for systems with other values of $\hat{q}$ and $\Mtotaltrue$.

\subsection{Mass ratio}
It is well known that the excitation of each HoM is sensitive to the mass ratio of a binary system.
In particular, Ref.~\citep{Blanchet:2008je} identified the mass ratio dependence of each mode amplitude in Post-Newtonian (PN) theory, which was later found to be in a good agreement with the full numerical relativity simulations~\cite{Borhanian:2019kxt}.
Broadly speaking, more HoMs are excited as the system is more asymmetric in component masses.
When the waveform model is sensitive to identify the presence (systems with more asymmetric masses) or the absence (systems with nearly equal masses), it can provide extra constraints on the parameter space along the degeneracy between inclination and distance.
Thus, the redshift uncertainties are generally smaller when one uses the HoM-waveform template in the inference, even if the true waveform does not contain substantial HoM contents.

\begin{figure}[!htb]
    \centering
    \includegraphics[width=0.9\columnwidth]{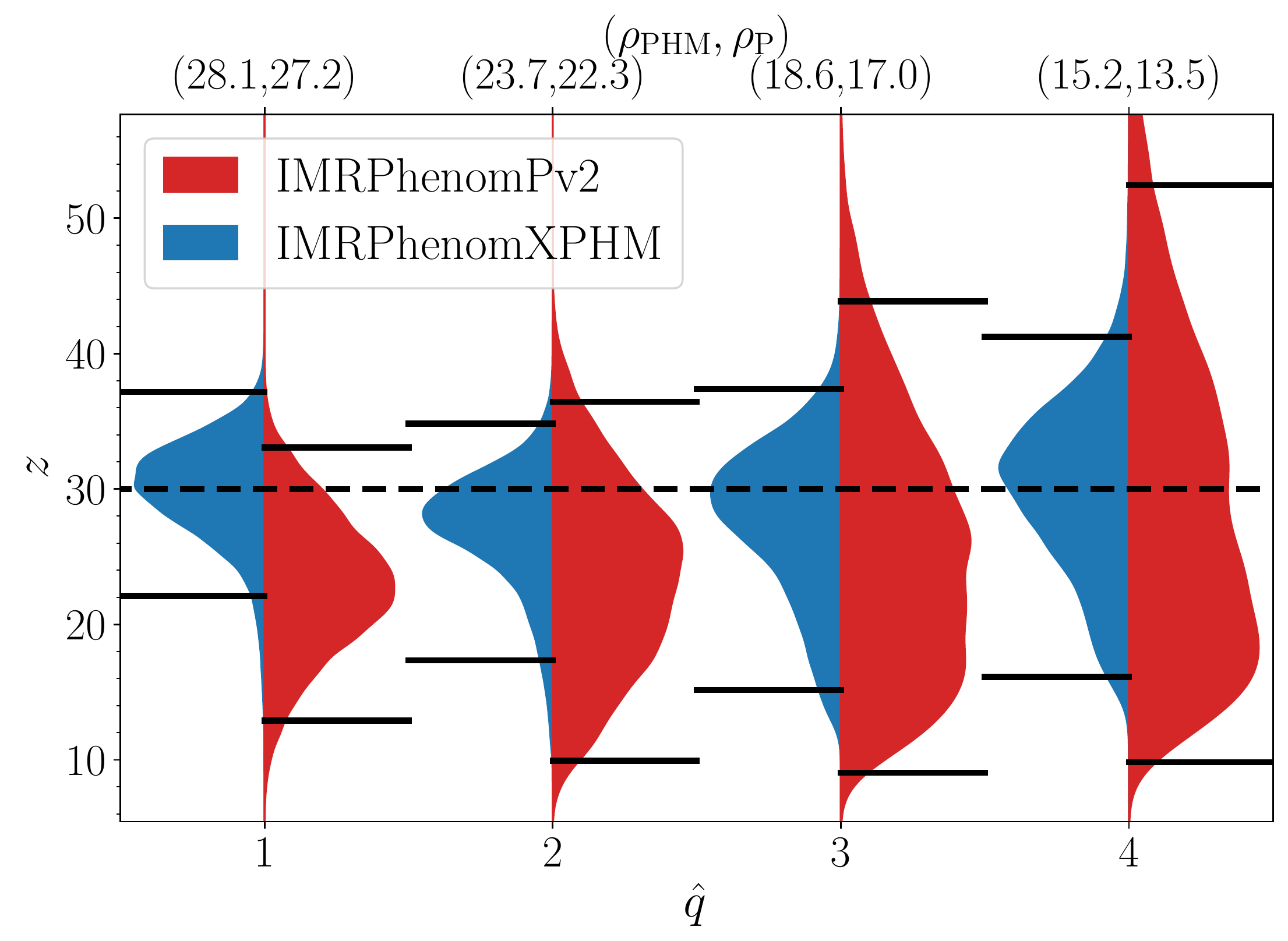}
    \caption{Redshift measurements for sources with $(\Mtotaltrue,\hat{z},\hat{\iota})=(40\msun,30,30^{\circ})$ at $\hat{q}=1,2,3$ and $4$. The format is the same as in Fig.~\ref{fig:zposterior_iota}.}
    \label{fig:zposterior_q}
\end{figure}
In Fig.~\ref{fig:zposterior_q}, we show the redshift posteriors obtained by the two waveform models for sources with $(\Mtotaltrue,\hat{z},\hat{\iota})=(40\msun,30,30^{\circ})$ at $\hat{q}=1,2,3$ and 4.
First, for all mass ratios, the redshift uncertainties obtained by \IMRXPHM decrease by $\sim20\%$ to $\sim 40\%$ when compared to those obtained by \IMRP.
Second, the scaling between the uncertainties and the SNRs in the \IMRXPHM is different from that in \IMRP.
While the increase in $q$ leads to an increase in the redshift uncertainty due to the decrease in the SNR, it also excites more the HoMs and enrich the structure in the \IMRXPHM waveform.
The additional HoM contents may provide more constraining power to break the distance-inclination degeneracy.
Hence, the redshift uncertainties obtained by \IMRXPHM do not increase with SNR as fast as those obtained by \IMRP.

\subsection{Total mass}
Unlike mass ratio and inclination, the total mass does not change the HoM contents, i.e., it simply scales the overall amplitudes and frequencies of all modes.
However, from the perspective of detection, it is a crucial parameter that determines the \textit{detectability} of HoMs, which then affects the ability of using HoMs to break the distance-inclination degeneracy.
HoMs have higher frequency, but weaker amplitudes, than the dominating $(2,2)$ mode.
If the power of all HoMs is much weaker than the given noise floor, only the $(2,2)$ mode is detectable and the distance-inclination degeneracy still remains, regardless of the presence of HoMs in the waveform model.

\begin{figure}[!htb]
    \centering
    \includegraphics[width=0.9\columnwidth]{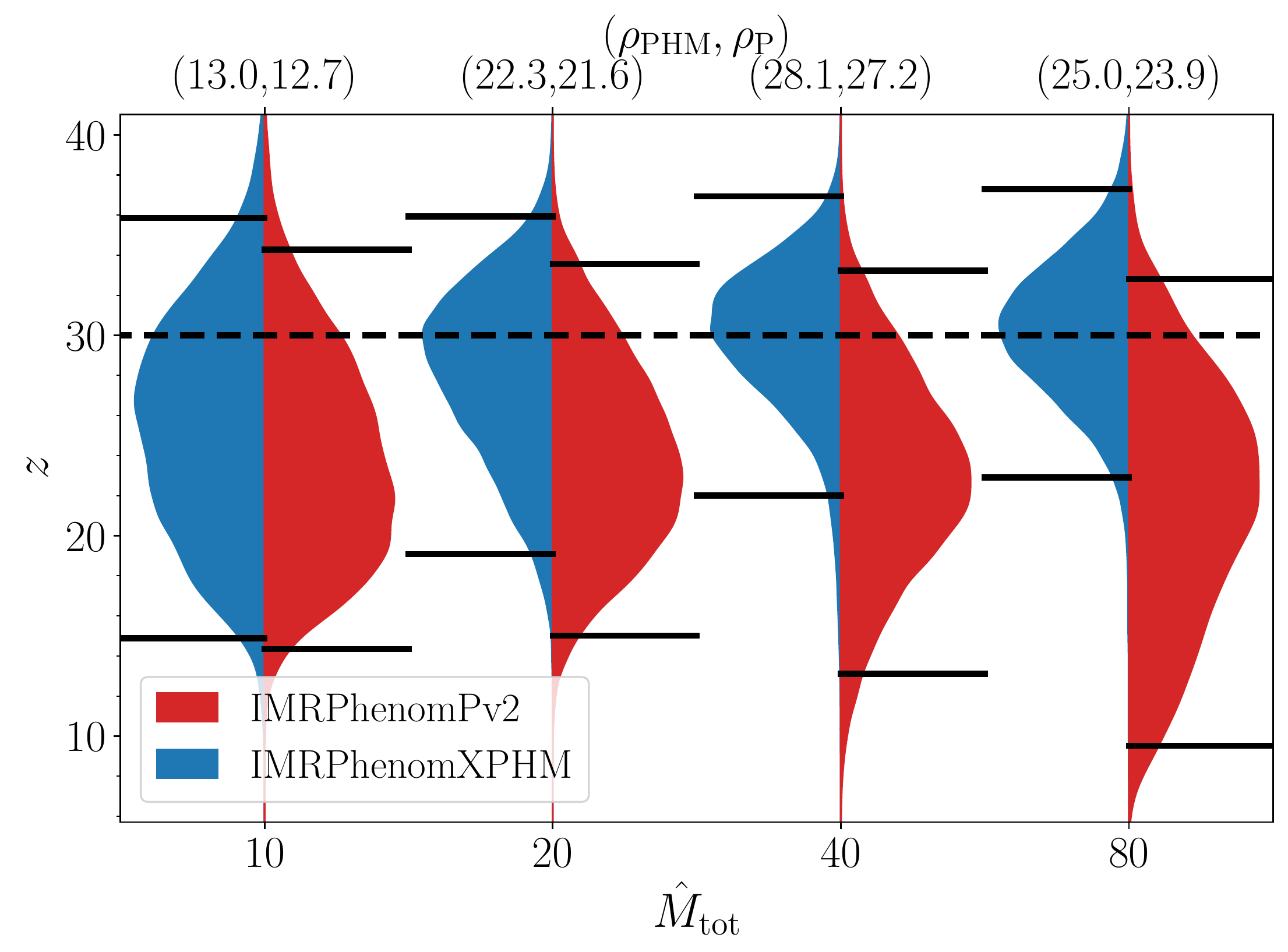}
    \caption{Redshift measurements for sources with $(\hat{q},\hat{z},\hat{\iota})=(1,30,30^{\circ})$ at $\Mtotaltrue=10,20,40$ and $80~\msun$. The format is the same as in Fig.~\ref{fig:zposterior_iota}.}
    \label{fig:zposterior_Mtot}
\end{figure}

As an example in Fig.~\ref{fig:zposterior_Mtot}, we show the posterior for the redshift obtained with the two waveform models for sources with $(\hat{q},\hat{z},\hat{\iota})=(1,30,30^{\circ})$ at $\Mtotaltrue=10,20,30$ and $40\msun$.
Indeed, the redshift uncertainty for $\Mtotaltrue=10\msun$ obtained by \IMRXPHM is comparable to those obtained by \IMRP.
On the other hand, as $\Mtotaltrue$ increases, the contribution of HoMs becomes significant and breaks the distance-inclination degeneracy.
The redshift uncertainties in the presence of HoMs can be reduced by $\sim 30-40\%$ when $\Mtotaltrue$ increases from $40$ to $80\msun$.
As $\Mtotaltrue$ increases further, the waveform drifts out of the sensitive frequency band, and the signal is not detectable.

\subsection{Redshift}\label{subsec:z_uncertainty_z}
\begin{figure}[!htb]
    \centering
    \includegraphics[width=0.9\columnwidth]{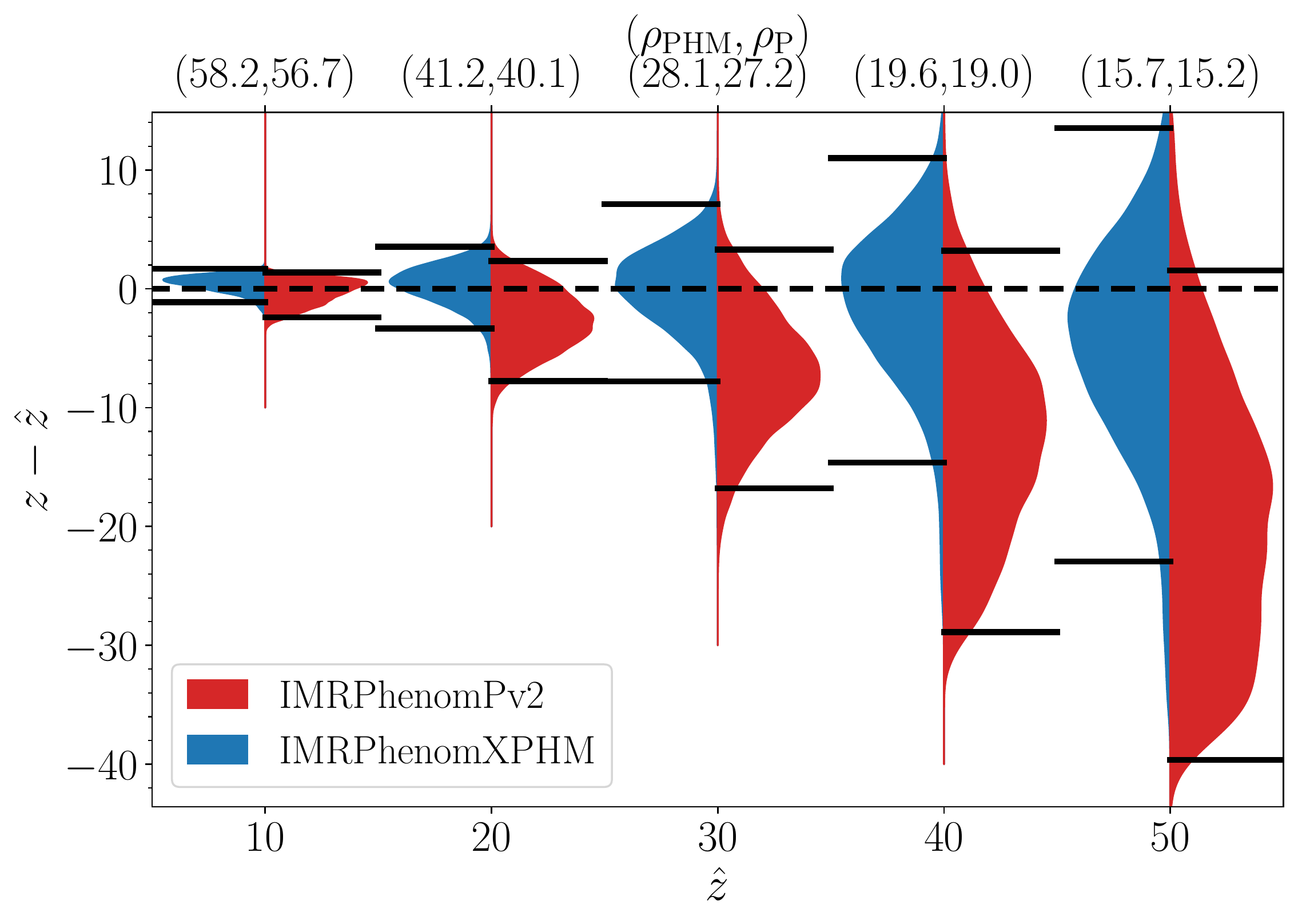}
    \caption{Redshift measurements for sources with $(\Mtotaltrue,\hat{q},\hat{\iota})=(40\msun,1,30^{\circ})$ at $\hat{z}=10,20,30,40$ and $50$.
    We offset the posterior to the true redshift $\hat{z}$. The format is the same as in Fig.~\ref{fig:zposterior_iota}.}
    \label{fig:zposterior_z}
\end{figure}

Similar to the total mass, the redshift only scales the overall amplitudes of all modes.
As the detector-frame total mass increases with the redshift, we expect the improvement on the redshift measurement to be larger at higher redshift.
On the other hand, the SNR decreases as the redshift increases, and the overall uncertainties increase in both waveform models.
These two trends can be seen in Fig.~\ref{fig:zposterior_z}, in which we compare the posteriors of redshifts obtained by the two waveform models for sources with $(\hat{M}_{\rm tot},\hat{q},\hat{\iota})=(40\msun,1,30^{\circ})$ at $\hat{z}=10,20,30,40$.
As the redshift increases, the posteriors of redshift in the cases of \IMRXPHM are centered at the true values, while those of \IMRP shift towards lower redshift.
This is a combined effect of the redshift prior and the distance-inclination degeneracy.
In the matter-dominated regime, $1\lesssim z\lesssim 1000$, this prior scales as $p_0(z) \sim (1+z)^{-5/2}$ and favors smaller redshift.
Due to the distance-inclination degeneracy in \IMRP, the posterior leans towards the region of larger inclination angle but smaller distance.
On the other hand, owing to the presence of HoMs, the posteriors obtained with \IMRXPHM do not suffer from this degeneracy and hence are less influenced by the inverse power-law scaling in $p_0(z)$.

\subsection{Correlation between distance, inclination and mass ratio}\label{subsec:q_iota_z}
\begin{figure}[h]
    \centering
    \includegraphics[width=0.9\columnwidth]{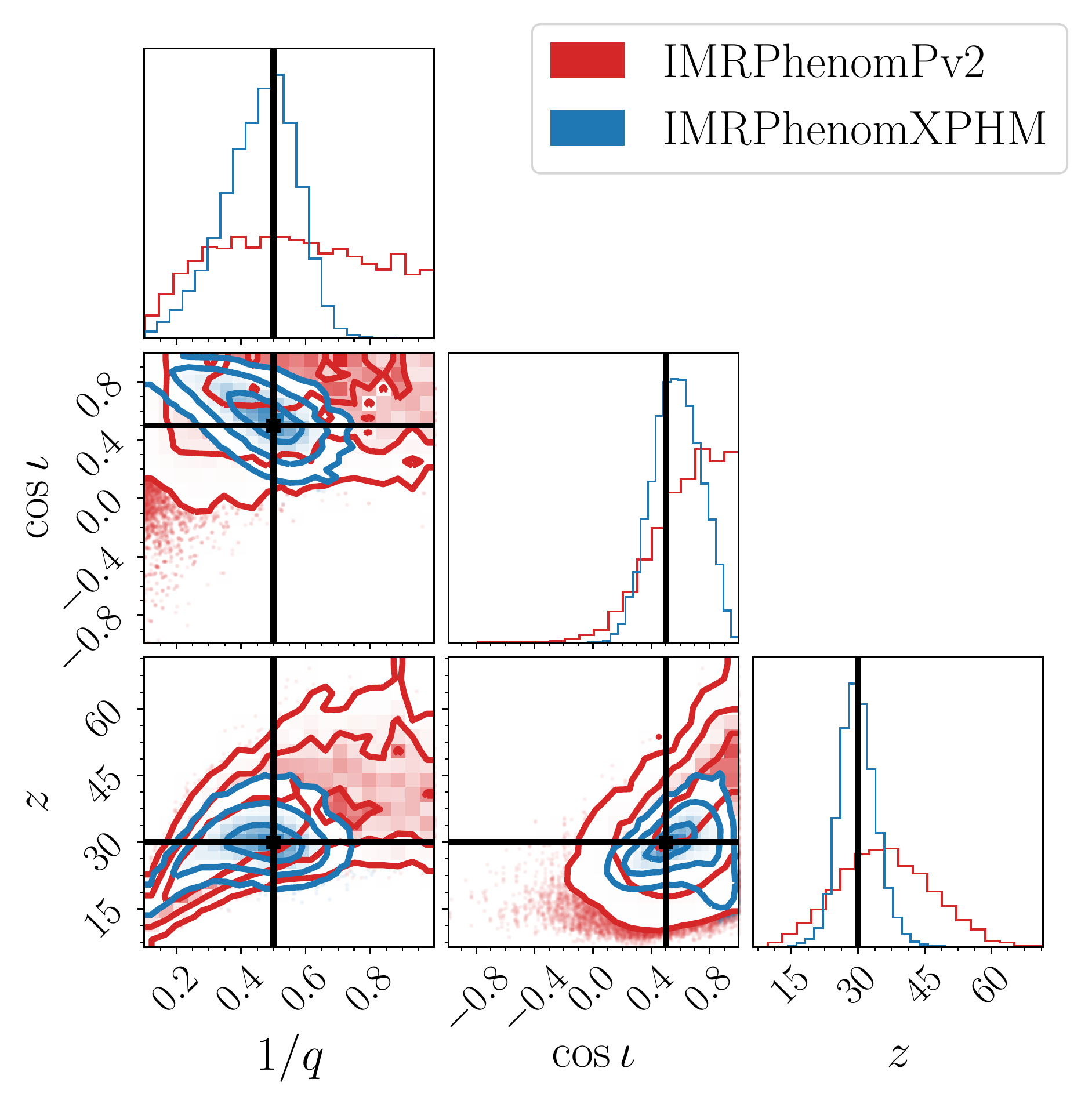}
    \caption{Comparison of the posteriors among $(1/q,\cos{\iota},z)$ obtained by \IMRP (red) and \IMRXPHM (blue) at $(\Mtotaltrue,\hat{z},\hat{q},\hat{\iota})=(40\msun,40,2,60^{\circ})$.
    The contours represent the boundaries of 68\%, 95\%, and 99.7\% credible regions.
    The true values are indicated by the black solid lines.
    }
    \label{fig:q_iota_z_corner_comp}
\end{figure}

We end this section by discussing how the presence of HoM aids the measurement in $(q,\iota,z)$.
We illustrate the two-dimensional (2D) posteriors among the pairs of $(q,\cos \iota,z)$, as shown in Fig.~\ref{fig:q_iota_z_corner_comp}.
In this example for the source with $(\Mtotaltrue,\hat{z},\hat{q},\hat{\iota})=(40\msun,40,2,60^{\circ})$, the 2D posteriors obtained by \IMRXPHM (blue) show drastically different behaviors when compared to those obtained by \IMRP (red).

First, the 2D contours, and hence the marginalized posteriors, are more localized in \IMRXPHM.
In particular, the marginalized posteriors of $1/q$ and $\cos \iota$ do not rail towards the prior edge at $q=\cos \iota = 1$.
Second, the contours in \IMRP posteriors have different structures than those in \IMRXPHM posteriors.
In the pair $(1/q,\cos \iota)$, the \IMRXPHM posterior shows an anti-correlation, but the \IMRP posterior is uncorrelated.
In the pair $(\cos \iota, z)$, the \IMRXPHM posterior partially follows the correlation as in \IMRP posterior for $\cos\iota\lesssim 0.6$, and becomes anti-correlated $\cos\iota\gtrsim 0.8$.
Similar situation happens in the pair $(1/q,z)$, in which the \IMRXPHM posterior becomes uncorrelated as $1/q$ increases to $\sim 0.4$.
These features emerge from the different dependence of $q$ and $\iota$ in each HoM.

\section{Mass and spin measurement}\label{sec:intrinsic}
In this section, we explore the impact of HoM on the improvement of measurements of masses and spins.
These intrinsic parameters can provide additional evidence for the (non)existence of PBHs, if they can be well measured.
For example, the spins of PBHs formed in the standard scenario from the collapse of density perturbations generated during inflation~\cite{Sasaki:2018dmp} are expected to be below the percent level, see Refs.~\cite{DeLuca:2019buf, Mirbabayi:2019uph}. The spins of PBHs are expected to remain negligible at $z\gtrsim30$ because there is not enough time for the BHs to gain angular momentum through accretion of surrounding materials. 
This prediction remain valid for $z\gtrsim10$ in the case of weak accretion~\cite{DeLuca:2020bjf}, which we will take as a benchmark scenario in the following.

The mass distribution of a population of PBHs depends on the properties of the collapsing perturbations. A large class of models predicts a distribution that can be approximated with a log-normal shape, whose central mass scale and width are model dependent and not currently observationally constrained.
If the masses of high-redshift BBHs are measured to be contradictory to the astrophysical predictions, these outliers can be smoking gun evidence for the existence of PBHs as well. The next section would be dedicated to the comparison with astrophysical populations.

We will first discuss the improvement on the measurement of the mass ratio $1/q$, then the source-frame primary mass $m_1$, and finally the effective spin $\chieff$.
\subsection{Mass ratio}
From the example of the $q-\iota-z$ joint measurement in Sec.~\ref{subsec:q_iota_z}, we expect that the presence of HoMs in the waveform model reduces the uncertainty in the mass ratio measurement.
Such improvement will also affect the measurement of the primary mass, which will be explored in the next subsection.
\begin{figure}[h]
    \centering
    \subfloat{\includegraphics[width=0.9\columnwidth]{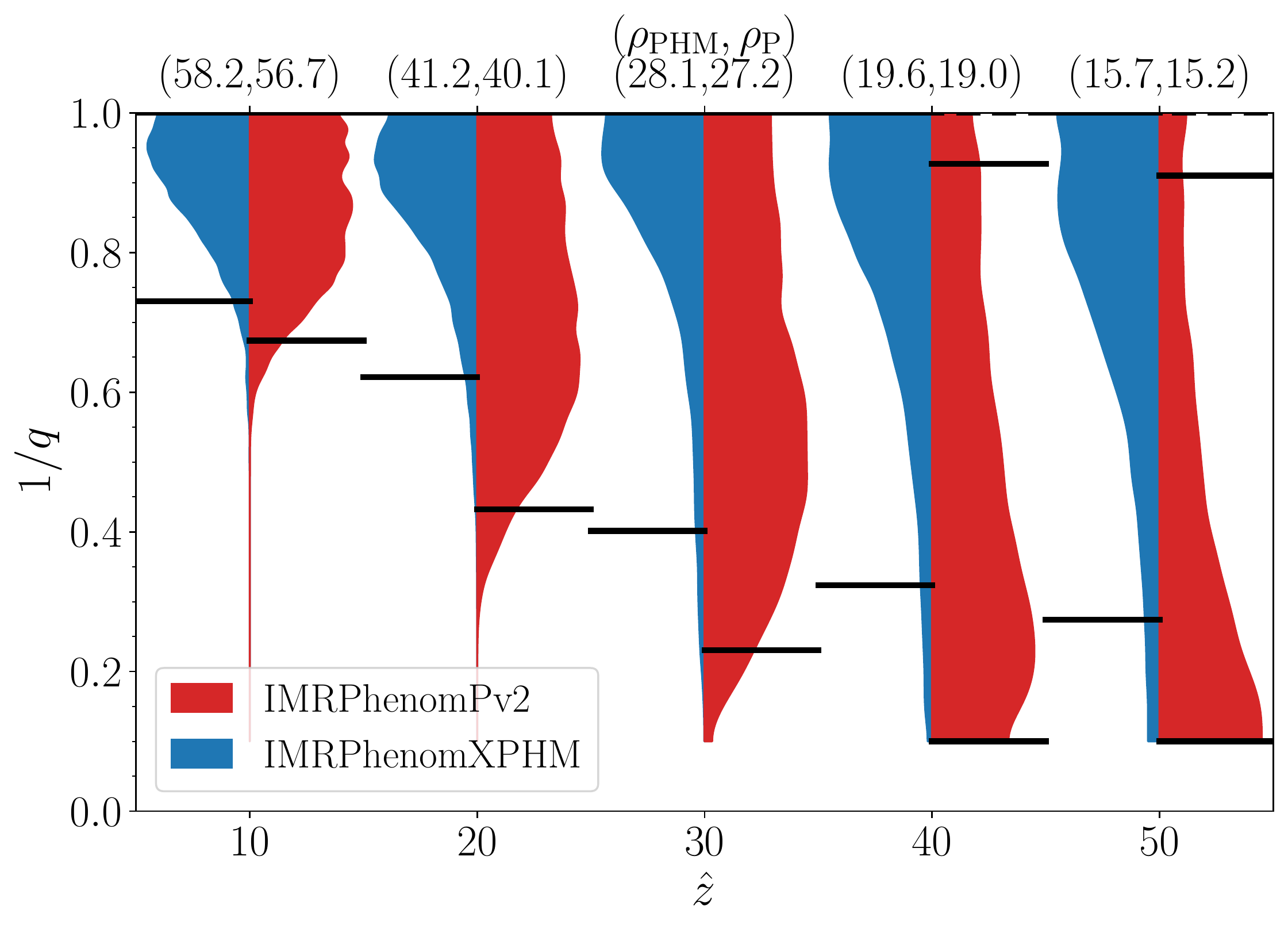}}
    \quad
    \subfloat{\includegraphics[width=0.9\columnwidth]{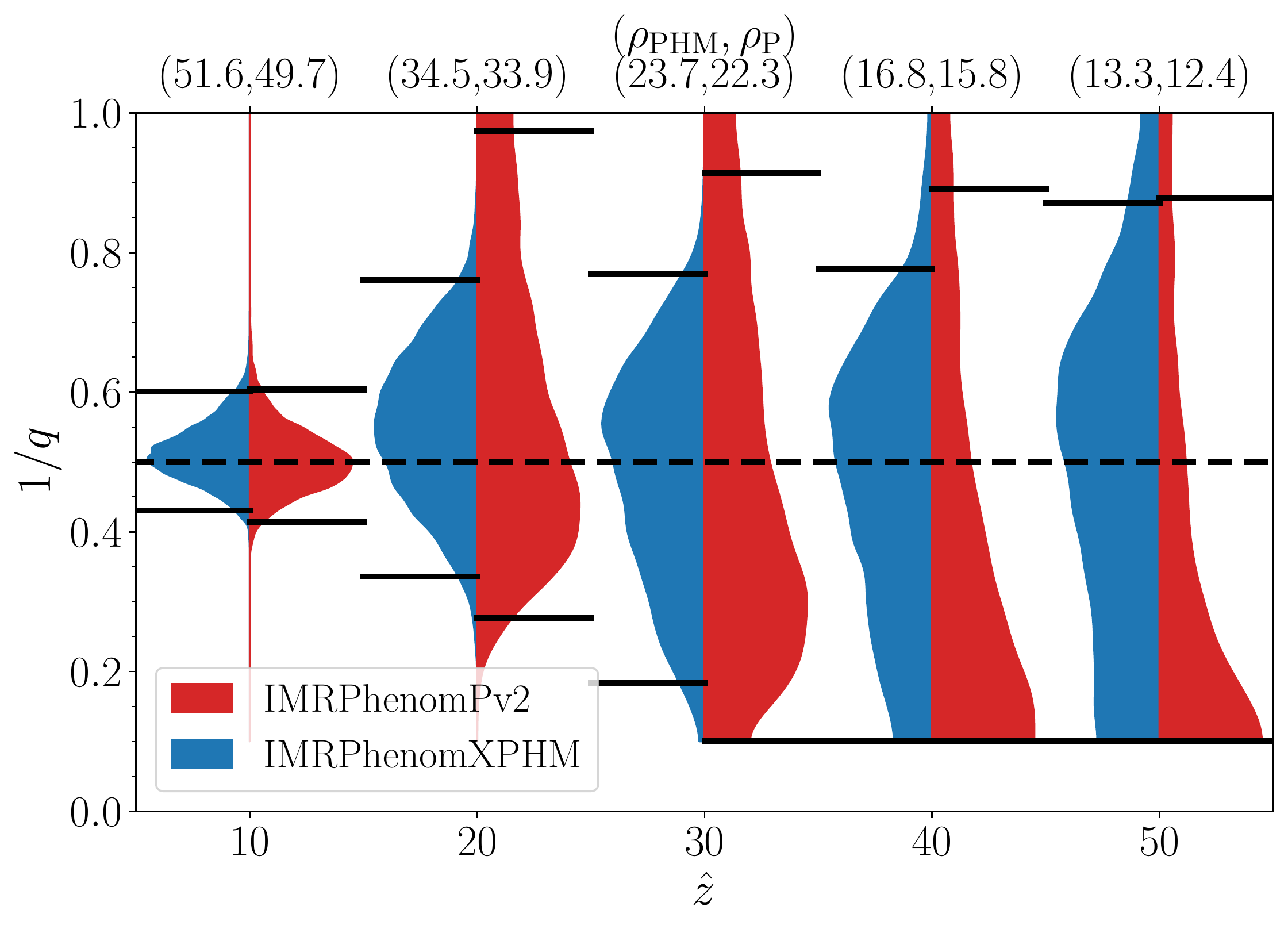}}
    \caption{Posteriors of mass ratio for sources with $(\Mtotaltrue,\hat{\iota})=(40\msun,30^{\circ})$ and $\hat{q}=1$ (top panel) or 2 (bottom panel) at $\hat{z}=10,20,30,40$ and 50, obtained by \IMRP (red) and \IMRXPHM (blue).
    Other plot settings are the same as in Fig.~\ref{fig:zposterior_iota}.}
    \label{fig:qposterior_z}
\end{figure}

In Fig.~\ref{fig:qposterior_z}, we compare the posteriors of mass ratio obtained by the two waveform models for sources with $(\Mtotaltrue,\hat{\iota})=(40\msun,30^{\circ})$ and $\hat{q}=1$ (top panel) or 2 (bottom panel) at $\hat{z}=10,20,30,40$ and 50.
Indeed, the uncertainties are generally smaller in the cases of \IMRXPHM.
The reduction of the uncertainties is more prominent when the redshift increases, because the detector-frame masses are larger and more HoMs are detectable.
In particular, at $\hat{z}\geq40$, the mass ratio is unconstrained in the cases of \IMRP, or even slightly biased away from $\hat{q}$.
Similar to the discussion in Sec.~\ref{subsec:z_uncertainty_z}, this is caused by the combined effect of the scaling in the redshift prior and the correlation between $(1/q,z)$ as shown in the \IMRP contours of Fig.~\ref{fig:q_iota_z_corner_comp}.
On the other hand, the posteriors in the cases of \IMRXPHM clearly show a peak around $\hat{q}$, i.e., consistent with the true value.

\subsection{Primary mass}\label{subsec:mass_uncertainty}
In GW astronomy, the mass parameters measured directly are the detector-frame chirp mass (total mass) if the waveform is dominated by the inspiral (merger-ringdown) phase within the sensitive frequency band.
Converting from the detector-frame chirp mass or total mass to the source-frame component masses, there are a factor of $(1+z)$ and a Jacobian term as a function of $q$.
Therefore, the improvement on both measurements of $q$ and $z$ propagate to the measurement of the source-frame primary mass.

\begin{figure}[h]
    \centering
    \includegraphics[width=0.9\columnwidth]{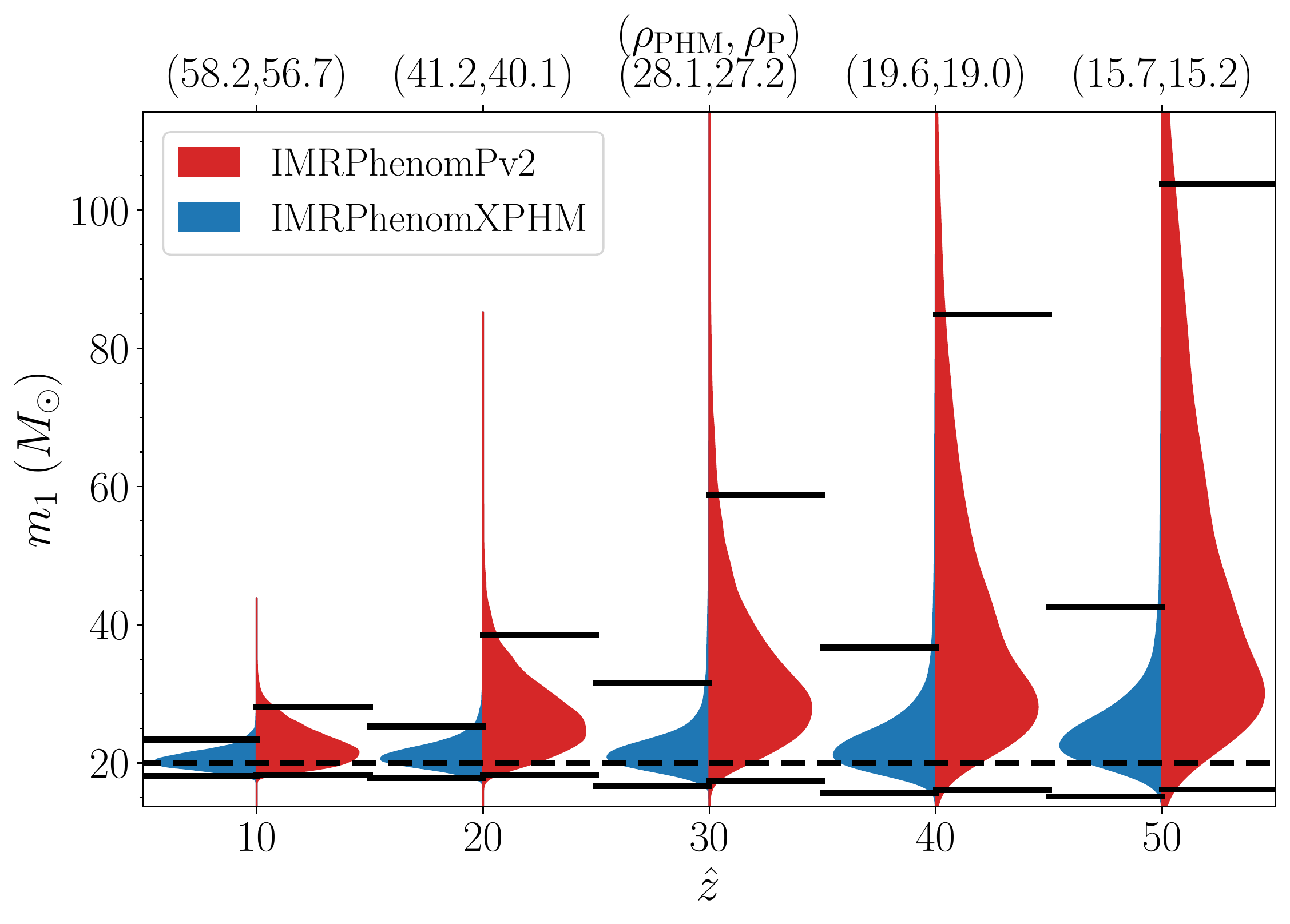}
    \quad
    \includegraphics[width=0.9\columnwidth]{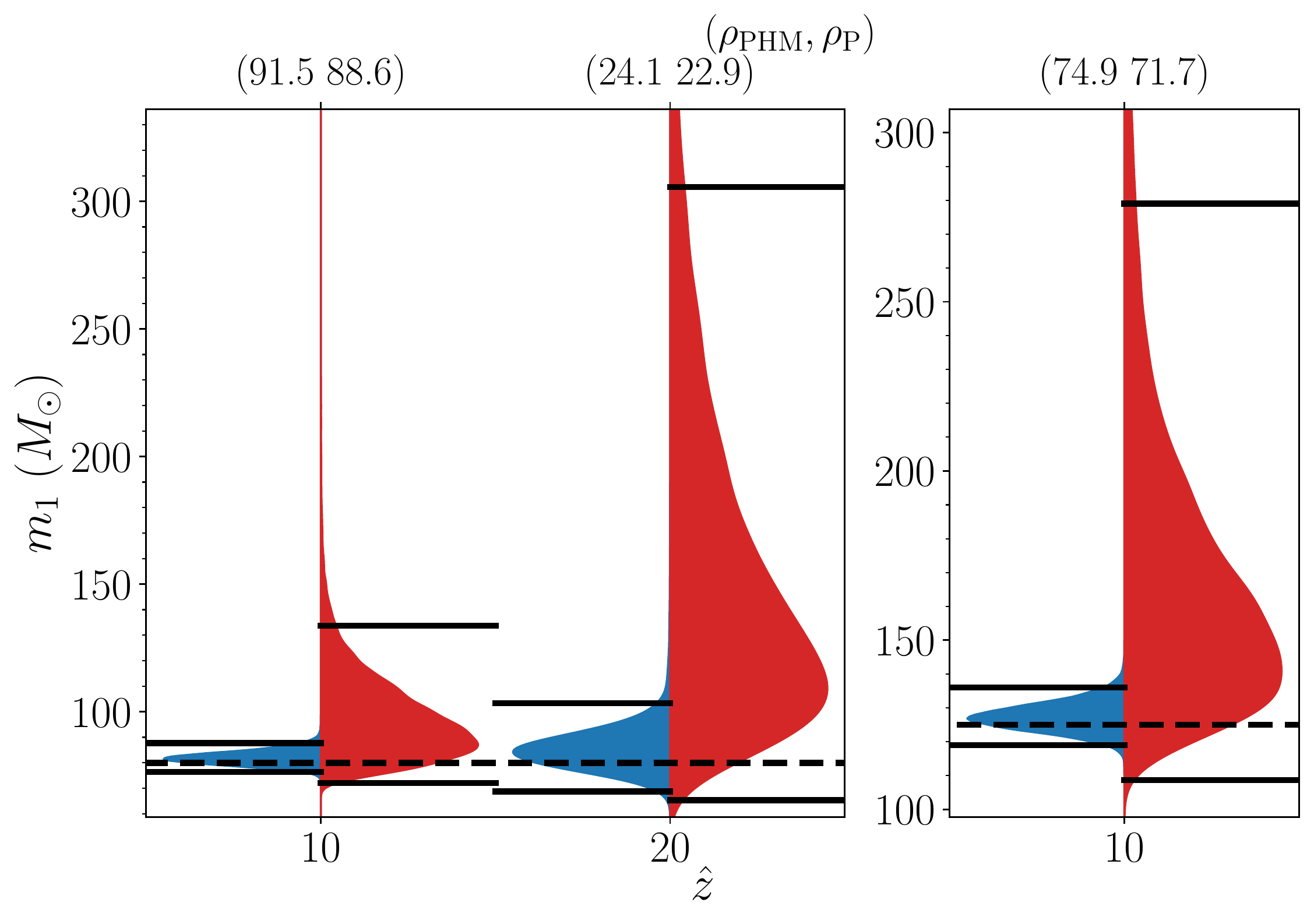}
    \caption{Posteriors of the source-frame primary mass for different sources obtained by \IMRP (red) and \IMRXPHM(blue).
    The parameters of the sources are $(\hat{m}_1,\hat{m}_2,\hat{\iota})=(20\msun,20\msun,30^{\circ})$ at $\hat{z}=10,20,30,40$ and 50 (upper panel), $(\hat{m}_1,\hat{m}_2,\hat{\iota})=(80\msun,80\msun,30^{\circ})$ at $\hat{z}=10$ and 20 (lower left panel), and $(\hat{m}_1,\hat{m}_2,\hat{\iota})=(125\msun,125\msun,30^{\circ})$ at $\hat{z}=10$ (lower right panel).
    Plot settings are the same as in Fig.~\ref{fig:zposterior_iota}.}
    \label{fig:m1srcposterior_z}
\end{figure}

In the upper panel of Fig.~\ref{fig:m1srcposterior_z}, we compare the posteriors of the source-frame primary mass obtained by the two waveform models for sources with $(\hat{m}_1,\hat{m}_2,\hat{\iota})=(20\msun,20\msun,30^{\circ})$ at $\hat{z}=10,20,30,40$ and 50.
The uncertainty improves by $\sim 40\%$ at $\hat{z}=10$ to $\sim 60\%$ at $\hat{z}=50$, owing to the presence of HoMs in the waveform which reduces the uncertainty in both the redshift and the mass ratio.

As the possible PBH masses are not restricted within the stellar mass region, we also show systems with a larger mass range, $\hat{m}_1=80$ and $125\msun$, but at smaller redshifts in the lower panel of Fig.~\ref{fig:m1srcposterior_z}.
The posterior of $m_1$ for the systems with $\hat{m}_1=80\msun$ at $\hat{z}=10$ (20) is well-constrained within a relative uncertainty of $\sim 10\%$ $(\sim 50\%)$ in the cases of \IMRXPHM, while those in the cases of \IMRP have $\geq5$ times larger uncertainties.
Similarly, for the system with $m_1=125\msun$, the relative uncertainty of $m_1$ is greatly reduced from $\sim 130\%$ (\IMRP) to $\sim 15\%$ (\IMRXPHM).

\subsection{Effective spin}
\begin{figure}[h]
    \centering
    \includegraphics[width=0.9\columnwidth]{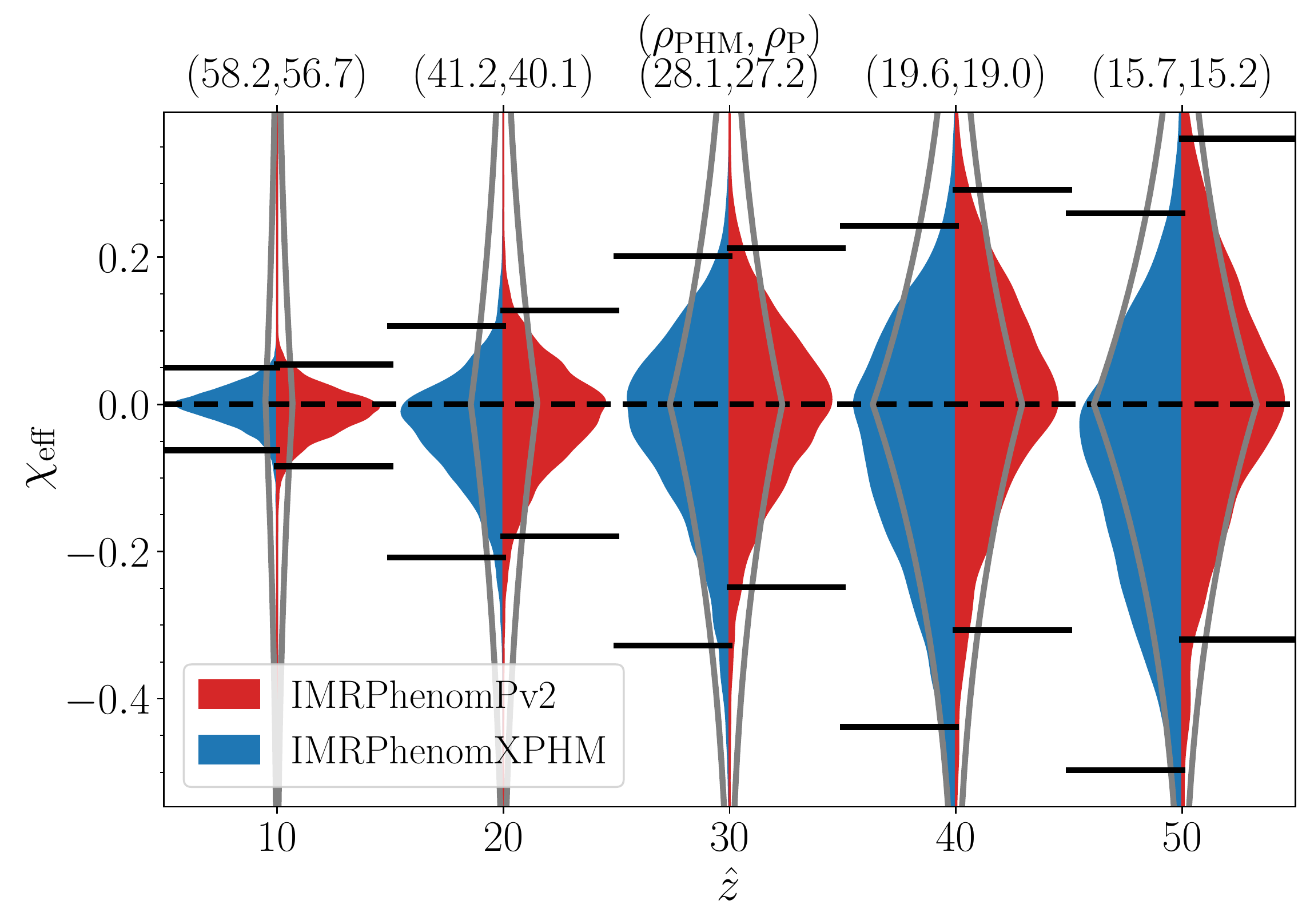}
    \caption{Posteriors of effective spin for zero-spin sources with $(\Mtotaltrue,\hat{q},\hat{\iota})=(40\msun,1,30^{\circ})$ at $\hat{z}=10,20,30,40$ and 50, obtained by \IMRP (red) and \IMRXPHM(blue).
    Grey solid lines indicate the prior on $\chieff$ sampled from the uniform prior on spin magnitudes, spin orientations, and $1/q$.
    For the sake of visualization, the areas of the violins are not normalized in common.
    Thus, the grey lines are only normalized to their corresponding violins and have different widths.
    Plot settings are the same as in Fig.~\ref{fig:zposterior_iota}.}
    \label{fig:chieffposterior_z}
\end{figure}

In this subsection, we focus on the weakly-accreting scenario~\cite{DeLuca:2020bjf} and demonstrate if the zero spin can be well measured, and whether the measurement can be improved by the presence of HoMs in the waveform model.
In Fig.~\ref{fig:chieffposterior_z}, we compare the posterior of $\chieff$ obtained by the two waveform models for zero-spin sources with $(\Mtotaltrue,\hat{q},\hat{\iota})=(40\msun,1,30^{\circ})$ at $\hat{z}=10,20,30,40$ and 50.
The uncertainties are comparable in both waveform models.
We note that the posteriors in the cases of \IMRXPHM are more asymmetric around 0, with larger probability masses in the negative $\chieff$ region, unlike the trends shown in Ref.~\cite{Ng:2018neg}.
Hence, we conclude that the presence of HoMs in the waveform model only changes the morphology of the likelihood function along $\chieff$, and does not improve the measurement of the spin parameters.

We also note that it is generally harder to measure the spins of the systems when the redshift increases.
This is because the spin effect first goes into the 1.5 PN \textit{inspiral} phase, but the signal drifts to the lower frequency and most of the SNR is dominated by the \textit{merger-ringdown} phase instead.

\section{Implications for the PBH detection}\label{sec:implications}
In the previous section, we have established that including HoMs in the waveform model improves the single measurement of the redshift and masses, while the uncertainty of effective spin remains unchanged.
In this section, we discuss how the single-event measurement of the redshift, primary mass and effective spin obtained by \IMRXPHM can be used to identify or infer the properties of PBHs.

\subsection{Redshift measurement}
The redshift may be one of the most significant parameter in the identification of PBHs.
This is because PBHs are formed much earlier than the astrophysical BHs.
While the time of birth of the first BH is still uncertain, theoretical and simulation studies suggest that the astrophysical epoch of BBHs is about $z\sim 30$~\cite{Kinugawa:2014zha,Kinugawa:2015nla,Hartwig:2016nde,Belczynski:2016ieo,Inayoshi:2017mrs,Liu:2020lmi,Liu:2020ufc,Kinugawa:2020ego,Tanikawa:2020cca}.

In Refs.~\cite{DeLuca:2021wjr,DeLuca:2021hde,Ng:2021sqn}, the critical redshift to approximate when the merger rate density of Pop~III BBHs turns off has been chosen to be $\zcrit=\zcritvalue$, i.e., the redshift region is \textit{fully primordial} above $\zcrit$.
With such definition, we showed that the relative abundance of PBH and Pop~III mergers can affect the significance of confirming the primordial origin of a single BBH detection.
The heuristic prior $\ptot$ based on the expected merger rate densities is then
\begin{align}
    \ptot&\left(z\vert \fcrit\right) \propto \nonumber \\
   & \left[\fcrit\frac{\npbh(z)}{\npbh(\zcrit)}+\frac{\nIII(z)}{\nIII(\zcrit)}\right]\frac{dV_c}{dz}\frac{1}{1+z},
\end{align}
where
\begin{align}\label{eq:popIII}
    \nIII(z) \propto
    \begin{cases}
     \frac{e^{\aIII(z-\zIII)}}{\bIII+\aIII e^{(\aIII+\bIII)(z-\zIII)}} &\mathrm{if }\,z<\zcrit \\
     0 &\mathrm{otherwise}
    \end{cases},
\end{align}
is the phenomenological fit to the Pop~III merger rate density based on the simulation study in Ref.~\cite{Belczynski:2016ieo,Ng:2020qpk}, with $(\aIII,\bIII,\zIII)=(0.66, 0.3, 11.6)$ from~\cite{Ng:2020qpk},
\begin{align}
    \npbh(z) \propto \left(\frac{t(z)}{t(0)}\right)^{-34/37},
\end{align}
is the analytic PBH merger rate density obtained from the dynamics of early PBH binary formation~\cite{Raidal:2017mfl,Chen:2018czv,Raidal:2018bbj,Chen:2019irf,DeLuca:2020qqa} (with $t(z)$ being the age of the Universe at $z$), and \begin{align}
\fcrit(\zcrit)\equiv \frac{\npbh(\zcrit)}{\nIII(\zcrit)}\nonumber
\end{align}
is the ratio between the two merger rate densities at $\zcrit$.

We also employed two statistical indicator in Ref.~\cite{Ng:2021sqn}.
The first indicator is the probability of primordial origin, $\Pp$ as the fraction of the redshift posterior with $z\geq \zcrit$,
\begin{align}
    \Pp (\zcrit\vert \fcrit) = \frac{1}{Z(\fcrit)}\int_{\zcrit}^{\infty} \frac{p\left(z\vert d\right)}{p_0(z)} \ptot\left(z\vert \fcrit\right) dz,
\end{align}
where $p\left(z\vert d\right)$ is the redshift posterior obtained with the default prior $p_0(z) \propto dV_c/dz/(1+z)$ (uniform source-frame rate density), and $Z(\fcrit)$ is the evidence of the merger rate density model $\ptot$ parameterized by $\fcrit$.
We found that the best system in our simulation set, $(\Mtotaltrue,\hat{q},\hat{\iota},\hat{z})=(40\msun,1,60^{\circ},40)$, i.e., the posterior mass has the most support at $z\geq\zcrit$, can only provides mild evidence for its primordial origin.
The second indicator is the Bayes factor between models differed by their priors evaluated at different $\fcrit$.
As the estimation of the Bayes factor is sensitive to the sampling algorithm and computationally expensive, we do not repeat the statistical analysis using the Bayes factor.
In the following, we relax the assumption of $\zcrit=\zcritvalue$, and explore how $\Pp$ varies as a function of $\zcrit\in[15,40]$.

\begin{figure}[h]
    \centering
    \includegraphics[width=0.9\columnwidth]{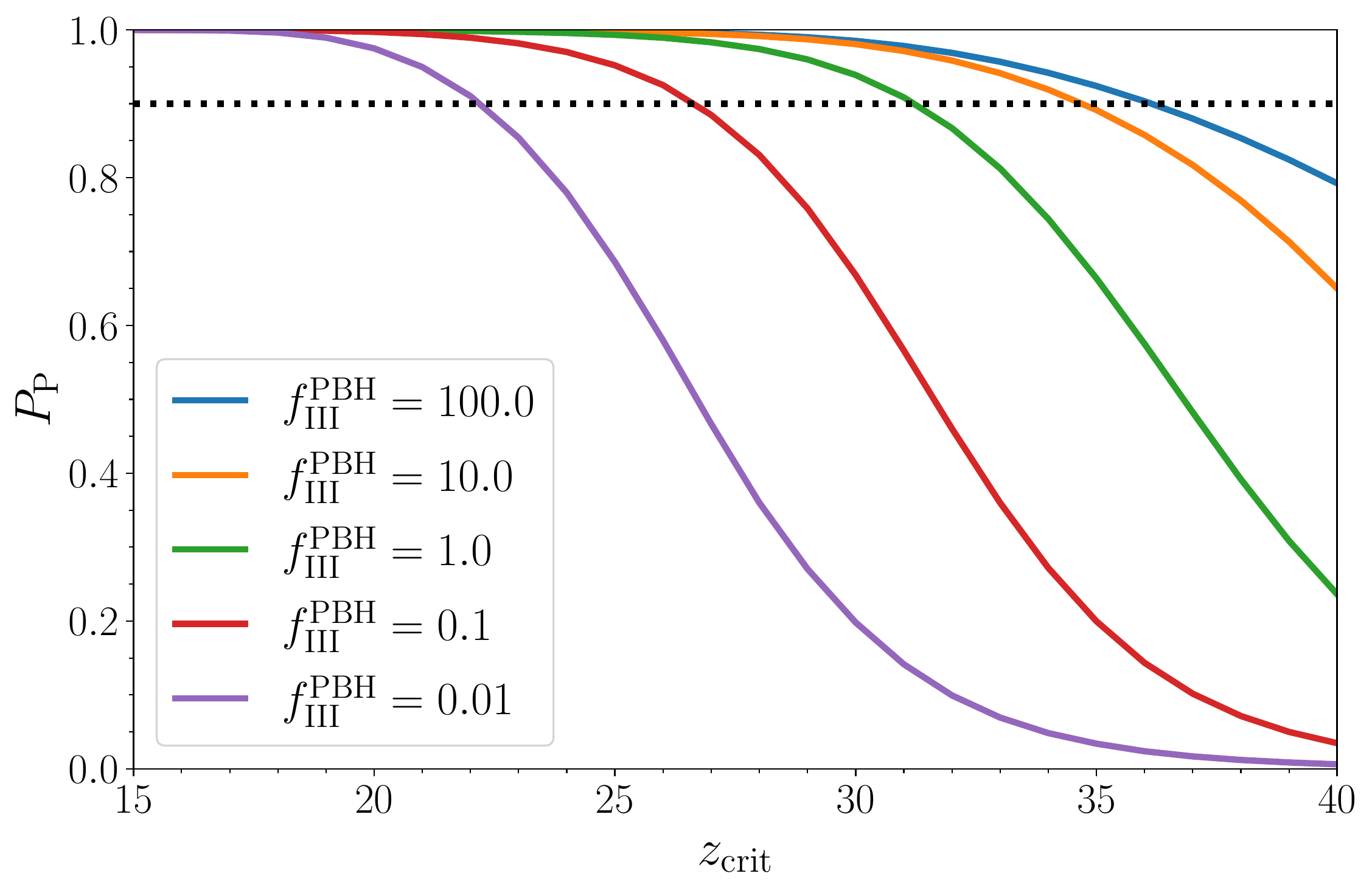}
    \caption{Probabilities of a primordial origin $\Pp$ as a function of $\zcrit$ for the system $(\Mtotaltrue,\hat{q},\hat{\iota},\hat{z})=(40\msun,1,60^{\circ},40)$ evaluated at $\fcrit=$100 (blue), 10 (orange), 1 (green), 0.1 (red) and 0.01 (purple).
    The calculations are based on the \IMRXPHM posteriors.}
    \label{fig:Pp_zcrit_fcrit}
\end{figure}

First, we revisit the system $(\Mtotaltrue,\hat{q},\hat{\iota},\hat{z})=(40\msun,1,60^{\circ},40)$ and show $\Pp(\zcrit)$ evaluated at $\fcrit=100,10,1,0.1$ and 0.01 in Fig.~\ref{fig:Pp_zcrit_fcrit}.
A smaller $\fcrit$ requires a lower $\zcrit$ to maintain $\Pp=0.9$ (black dotted line).
This is expected since the decrease in the prior volume of $z\geq \zcrit$ due to a smaller $\fcrit$ can be compensated by lowering $\zcrit$ for a larger primordial region.
In this example, even if $\fcrit=0.01$ at $\zcrit\sim 23$, the redshift measurement still allows for $\Pp=0.9$.

\begin{figure*}
    \centering
    \subfloat[$(\Mtotaltrue,\hat{z},\hat{q})=(40\msun,30,1)$\label{subfig:Pp_iota}]{\includegraphics[width=0.45\textwidth]{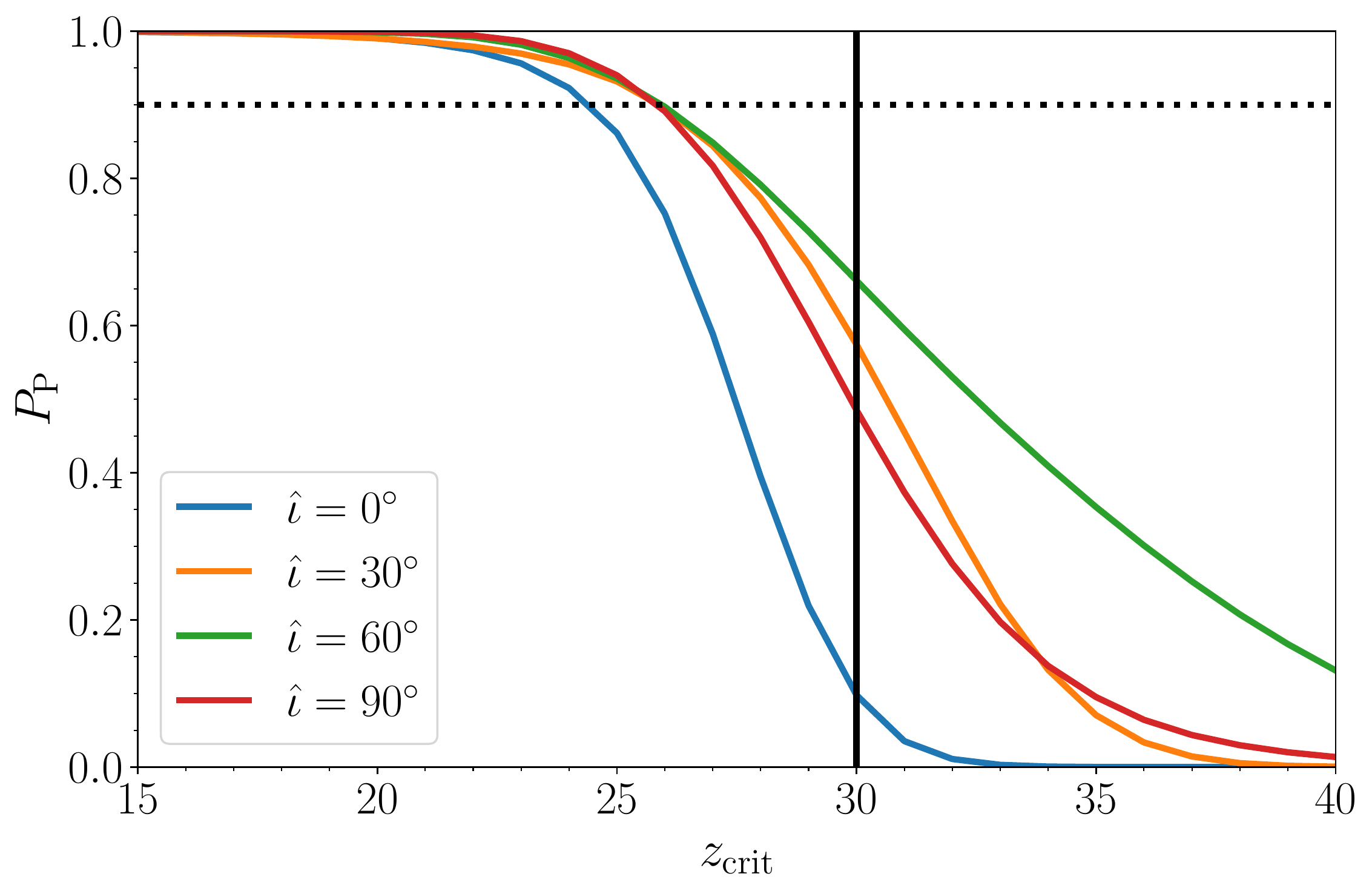}}
    \hfill
    \subfloat[$(\Mtotaltrue,\hat{\iota},\hat{z})=(40\msun,30^{\circ},30)$\label{subfig:Pp_q}]{\includegraphics[width=0.45\textwidth]{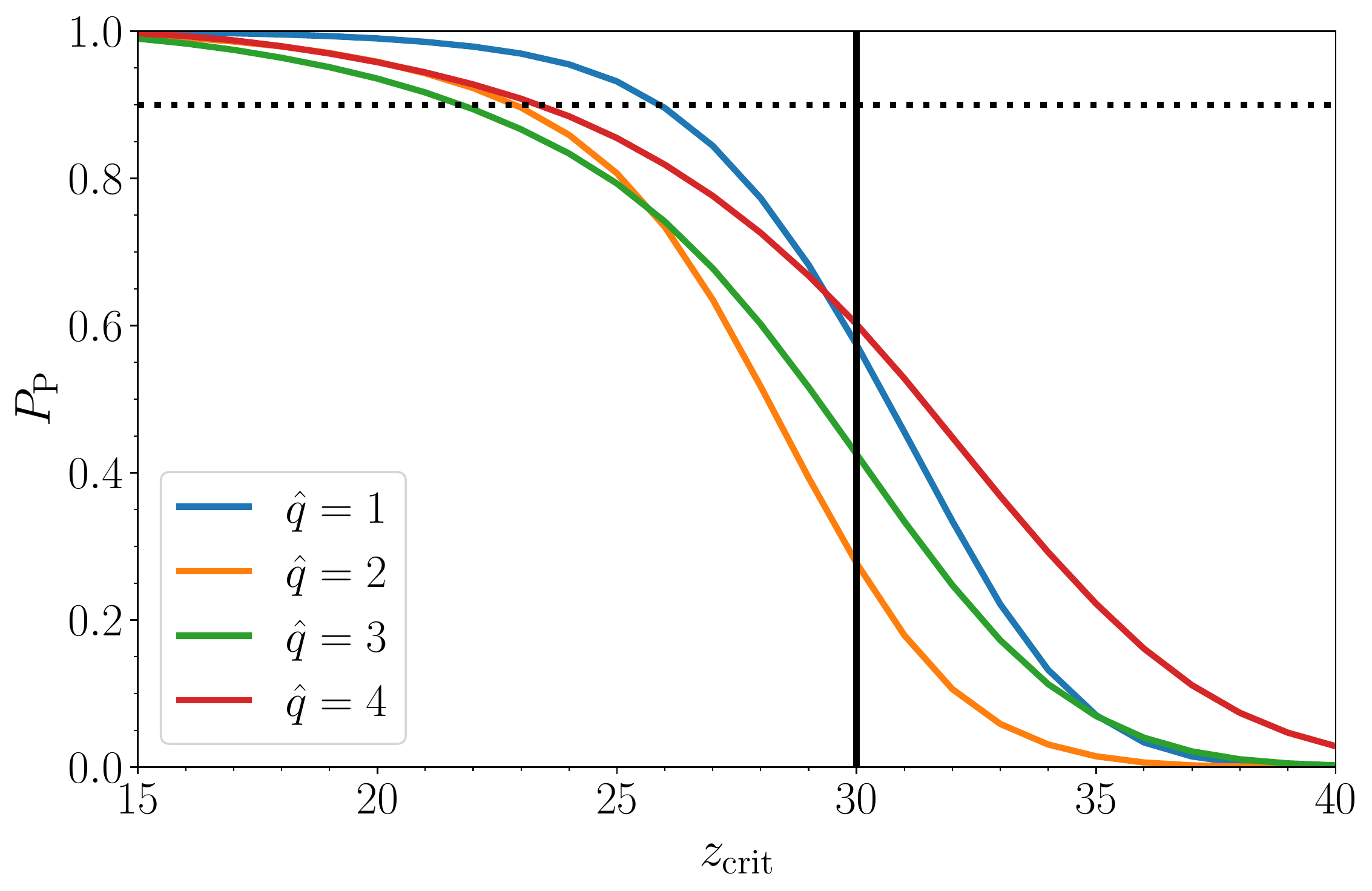}}
    \quad
    \subfloat[$(\hat{q},\hat{\iota},\hat{z})=(1,30^{\circ},30)$\label{subfig:Pp_mtotal}]{\includegraphics[width=0.45\textwidth]{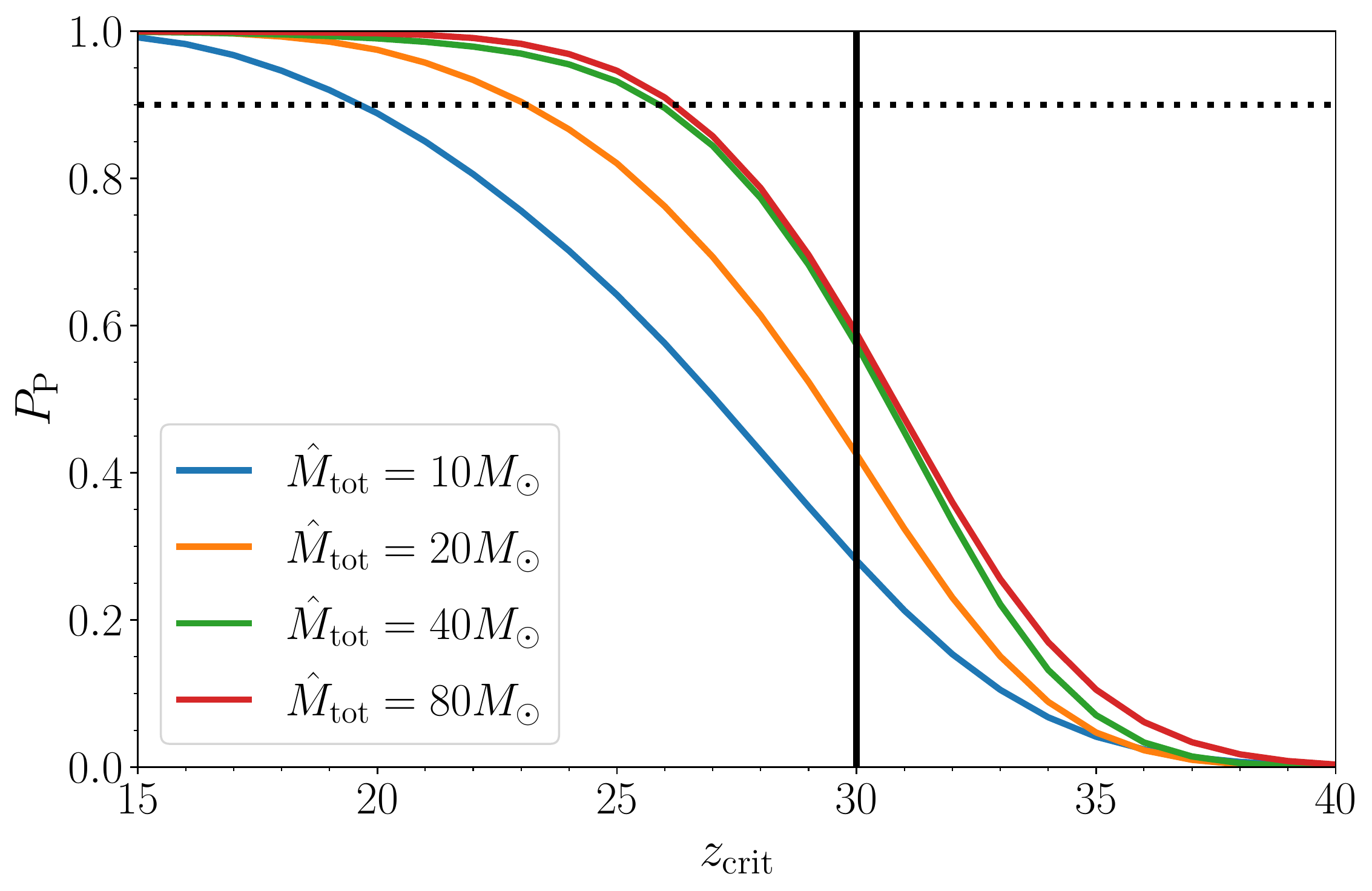}}
    \hfill
    \subfloat[$(\Mtotaltrue,\hat{q},\hat{\iota})=(40\msun,1,30^{\circ})$\label{subfig:Pp_z}]{\includegraphics[width=0.45\textwidth]{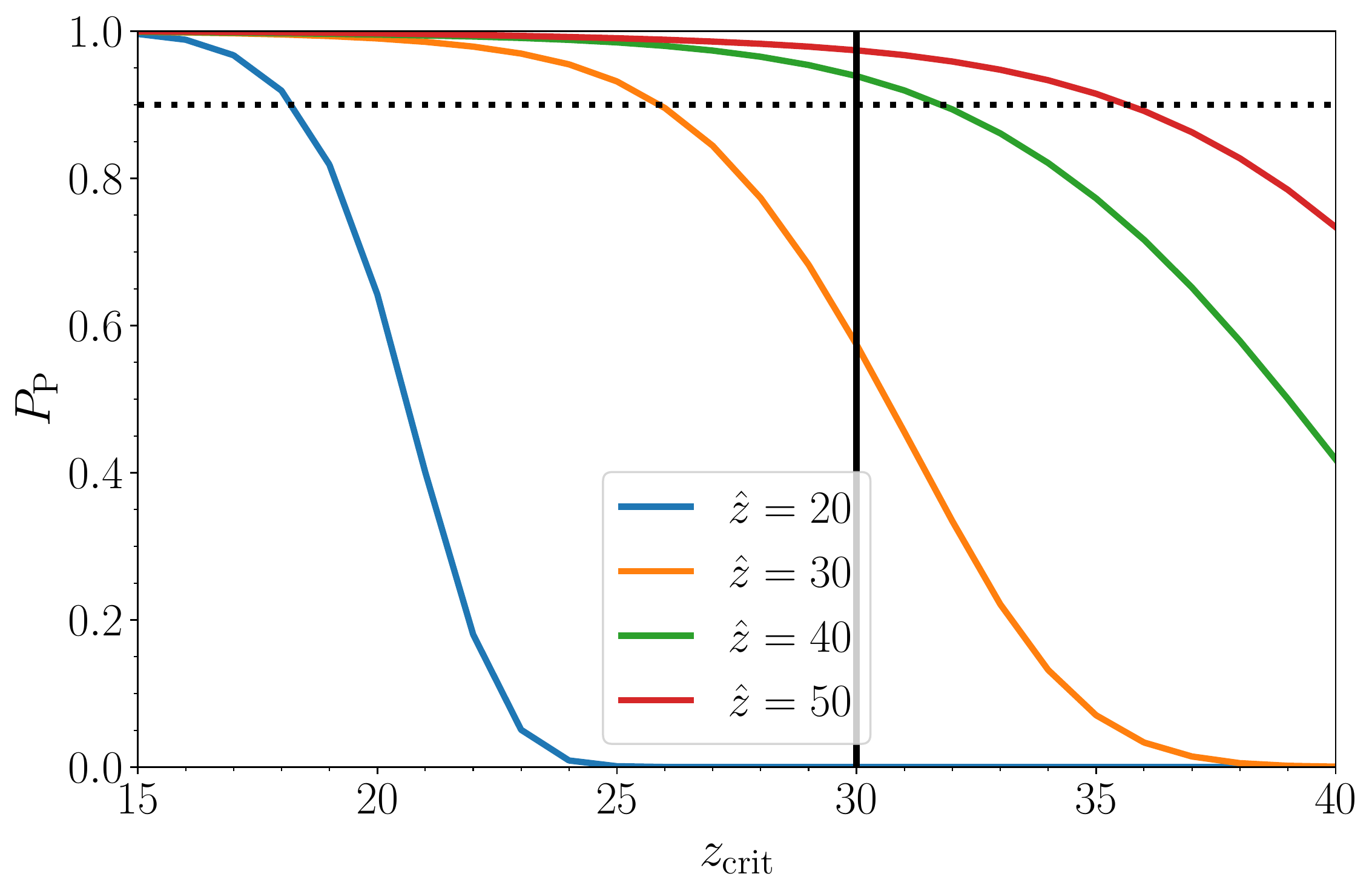}}
    \caption{Probabilities of a primordial origin $\Pp$ as a function of $\zcrit$ for the systems with (a) $(\Mtotaltrue,\hat{q},\hat{z})=(40\msun,1,30)$ and $\iota=0^{\circ}$ (blue), $30^{\circ}$ (orange), $60^{\circ}$ (green) and $90^{\circ}$ (red);
    (b) $(\Mtotaltrue,\hat{\iota},\hat{z})=(40\msun,30^{\circ},30)$ and $\hat{q}=$1 (blue), 2 (orange), 3 (green) and 4 (red);
    (c) $(\hat{q},\hat{\iota},\hat{z})=(1,30^{\circ},30)$ and $\Mtotaltrue=$10 (blue), 20 (orange), 40 (green) and 80$\msun$ (red);
    as well as (d) $(\Mtotaltrue,\hat{q},\hat{\iota})=(40\msun,1,30^{\circ})$ and $z=20$ (blue), 30 (orange), 40 (green) and 50 (red), all evaluated at $\fcrit=100$.
    The calculations are based on the \IMRXPHM posteriors.
    The choice of $\zcrit=30$ is marked by the black solid line.}
    \label{fig:Pp_zcrit_params}
\end{figure*}

Next, we choose the optimistic scenario, $\fcrit=100$, and illustrate how $\Pp(\zcrit)$ differs for various sets of source parameters.
We \textit{pivot} the system $(\Mtotaltrue,\hat{q},\hat{\iota},\hat{z})=(40\msun,1,30^{\circ},30)$, and vary the values of these four parameters one by one, as shown in each panel of Fig.~\ref{fig:Pp_zcrit_params}.
The intersection of the black dotted line and each colored line indicates the maximum $\zcrit$ to reach $\Pp=0.9$ for each system.
Most of the systems have the values of maximum $\zcrit$ above 20, except for the system at $\hat{z}=20$ which has a maximum $\zcrit \sim 18$.
The shift of maximum $\zcrit$ is a direct result of the change of the redshift uncertainties due to the change of $\hat{\iota}, \hat{q}, \Mtotaltrue$ and $\hat{z}$ as shown in Figs.~\ref{fig:zposterior_iota},~\ref{fig:zposterior_q},~\ref{fig:zposterior_Mtot}~and~\ref{fig:zposterior_z}, respectively.

\subsection{Mass measurement}
The improvement on the primary mass measurement as discussed in Sec.~\ref{subsec:mass_uncertainty} certainly helps the statistical inference of the PBH mass spectrum (if they exist within the detectable mass range).
Roughly speaking, the uncertainties of the population properties (which can be the median mass and variance of a log-normal mass spectrum in the context of PBHs) scale with the uncertainties of the single measurement, and inversely with the square root of the number of detections.
A factor of two improvement on the mass measurement (as seen in Fig.~\ref{fig:m1srcposterior_z}) translates to a factor of four reduction of the required number of events to reach the same statistical uncertainties of the population properties.

Moreover, the improvement on the mass measurement may also increase the statistical power for the identification of PBHs whose masses are outside the astrophysically allowed values.
Astrophysical BBHs originated from Pop I/II stars are not expected to have masses lying in the pair-instability supernova (PISN) mass gap $(\sim 50-130\msun)$~\cite{Belczynski:2016jno,Spera:2017fyx,Woosley:2016hmi}, unless the mergers consist of remnants of previous mergers in dense stellar clusters~\cite{Rodriguez:2017pec} or gas-rich environment~\cite{Yang:2019cbr}.
Recent studies of BBHs originated from Pop~III stars suggest that the mass gap may be narrowed to $\sim 100-130\msun$~\cite{Tanikawa:2021qqi}.
In both cases, a good measurement of $m_1$ is valuable in both measuring the mass spectra of Pop~III and PBH mergers, or serves as a possible indicator for the existence of PBHs provided that the Pop~III mergers cannot fill up the mass gap efficiently at high redshifts.
Given the model uncertainties in Pop~III mergers, let us stick with the ``standard'' PISN mass gap scenario: $\sim 50-130\msun$.
In Fig.~\ref{fig:m1srcposterior_z}, the mass uncertainties of the in-gap systems obtained by \IMRXPHM lie within the mass gap, while those obtained by \IMRP do not.
This shows the importance of HoMs for identifying outliers that are inconsistent with the astrophysical predictions, such as the existence of PBHs.

\subsection{Spin measurement}
The spin spectrum of the Pop~III BBHs is still highly uncertain, since both the initial conditions of the Pop~III stars and the binary evolution under very low metallicity are yet to be understood.
Pop~III BBHs may have non-negligible aligned spins, i.e., the majority of the BBHs have $\chieff\geq0$~\cite{Tanikawa:2021qqi}.
On the other hand, PBH binaries are dynamically formed.
Even if each component PBH acquires a significant spin through accretion, the distribution of $\chieff$ is expected to be symmetric around 0~\cite{DeLuca:2020bjf}.
This may be another signature to distinguish between Pop~III and primordial BBH population at high redshift.
In fact, such method has been proposed to measure the branching ratio between the formation channels of the dynamical capture and the isolated binary evolution using the LIGO/Virgo BBHs within $z<1$~\cite{Vitale:2015tea,Farr:2017gtv,GWTC2rate}.

As seen in Fig.~\ref{fig:chieffposterior_z}, including HoMs does not improve the effective spin measurement.
Also, the typical uncertainties of these high-redshift BBHs are comparable to those of the current detections~\cite{GWTC1,GWTC2,GWTC3}.
Therefore, the required number of detections needed for distingushing the formation channels may be $\mathcal{O}(100)$, similar to that predicted in previous studies~\cite{Farr:2017gtv,Vitale:2015tea}.

\section{Discussions}
In this work, we have simulated BBHs merging at $z\geq10$ with different mass ratios, total masses, and inclinations.
Performing full Bayesian parameter estimation with a waveform model that includes HoMs and spin precession, we quantified the uncertainties in redshifts, mass ratios, source-frame primary masses, and effective spins, as measured by a network of next-generation ground-based GW detectors.
We compared the uncertainties obtained by a waveform model without HoMs, and found that adding HoMs to the waveform model can improve the measurements of redshifts and component masses by up to a factor of 5.
Such improvements originate from the different dependence of mass ratio and inclination in the amplitude of each HoM, which helps breaking the $(2,2)$ distance-inclination degeneracy and constraining the mass ratio.
Generally speaking, BBHs with more asymmetric mass ratios, more inclined orbit, or larger masses in the detector frame are more benefited by the HoMs.
We note that the improvements are sensitive to the exact configuration of the BBH, which needs to generate large enough amplitudes of HoMs in order to break the degeneracy~\cite{Mills:2020thr}.
We also showed that including HoMs in the waveform models has no effect on the spin measurements for zero-spin sources.

We note that waveform systematics due to different descriptions of the spin precession dynamics are still present in the waveform families without HoMs, namely \IMRP and \IMRXP~\cite{Pratten:2020fqn,Garcia-Quiros:2020qpx}.
However, the improvements shown in this work are driven by the inclusion of HoMs, which breaks the distance-inclination degeneracy (see Fig.~\ref{fig.wfs}), not by the detailed description of the spin dynamics.
Therefore, we expect our results would remain unchanged if the systems have non-negligible spins.
Given the improved uncertainties in redshifts, we revisit the evaluation of primordial probabilities, which depend on both the critical redshift that defines the epoch of astrophysical BBHs, and the relative abundance characterized by the rate density ratio between Pop~III and PBH mergers at the critical redshift.
With the best redshift measurements in our simulation set, one may ensure the primordial probability larger than $90\%$ for a relative abundance $\fcrit(\zcrit)\geq0.01$, if the critical redshift is $\gtrsim 20$.
We also explored how BBHs with different parameters react to the calculation of primordial probabilities.
Assuming a large relative abundance $\fcrit(\zcrit)=100$, the minimum $\zcrit$ to achieve a primordial probability of 90\% is $\gtrsim 20$, depending on the true redshifts of the systems mostly.

The improvements on the measurement of masses enable a better inference of the mass spectra of both the Pop~III and PBH mergers.
While the measurements of effective spins remain mostly unchanged, the uncertainties are comparable to those of the current measurements.
One may borrow the method of using the distribution of effective spins to distinguish these two high-redshift BBHs based on different formation scenarios as suggested by the existing literature.
In any case, we expect the mass and spin measurements to provide additional evidence for identifying PBHs using high-redshift GW observations.
We will investigate this avenue as a future work.

\section{Acknowledgments} 
We would like to thank Valerio De Luca for suggestions and comments.
KKYN and SV are supported by the NSF through the award PHY-1836814. SC is supported by the Undergraduate Research Opportunities Program of Massachusetts Institute of Technology.
The work of MM is supported by the Swiss National Science Foundation and by the SwissMap National Center for Competence in Research. MB acknowledges support from the European Union’s Horizon 2020 Programme under the AHEAD2020 project (grant agreement n. 871158). BSS is supported in part by NSF Grant No. PHY-2207638, AST-2006384 and PHY- 2012083. SB is supported by NSF Grant No.PHY-1836779. BG is supported by the Italian Ministry of Education, University and Research within the PRIN 2017 Research Program Framework, n. 2017SYRTCN.
A.R. are supported by the Swiss National Science Foundation 
(SNSF), project {\sl The Non-Gaussian Universe and Cosmological Symmetries}, project number: 200020-178787.
G.F. acknowledges financial support provided under the European Union's H2020 ERC, Starting Grant agreement no.~DarkGRA--757480, and under the MIUR PRIN and FARE programmes (GW-NEXT, CUP:~B84I20000100001), and support from the Amaldi Research Center funded by the MIUR program ``Dipartimento di Eccellenza" (CUP:~B81I18001170001)
and H2020-MSCA-RISE-2020 GRU.

\bibliography{pbh}

\end{document}